\documentclass[11pt,onecolumn]{IEEEtran}
\usepackage{amsmath,amssymb,eucal,graphicx}
\usepackage{amsfonts}
\usepackage{epsfig}
\usepackage{calc}
\usepackage{subfigure}

\usepackage{latexsym}
\usepackage{verbatim}

\newtheorem{thm}{Theorem}
\newtheorem{lemma}{Lemma}
\newtheorem{cor}{Corollary}[section]

\newtheorem{remark}{\indent \bf Remark}[section]

\newcommand{\define}{\stackrel{\Delta}{=}}

\renewcommand{\QED}{\QEDopen}

\def\LSB{\left[}        
\def\RSB{\right]}       
\def\LB{\left(}         
\def\RB{\right)}        

\usepackage[usenames]{color}
\usepackage[normalem]{ulem}

\long\def\comment#1{}

\newfont{\bb}{msbm10 scaled 1000}
\newcommand{\CC}{\mbox{\bb C}}

\newcommand{\EE}{\mbox{\bb E}}

\newfont{\bbsmall}{msbm10 scaled 700}

\newcommand{\EEsmall}{\mbox{\bbsmall E}}

\newcommand{\av}{{\bf a}}
\newcommand{\bv}{{\bf b}}

\newcommand{\ev}{{\bf e}}

\newcommand{\gv}{{\bf g}}
\newcommand{\hv}{{\bf h}}

\newcommand{\nv}{{\bf n}}

\newcommand{\pv}{{\bf p}}

\newcommand{\sv}{{\bf s}}

\newcommand{\uv}{{\bf u}}
\newcommand{\wv}{{\bf w}}
\newcommand{\vv}{{\bf v}}
\newcommand{\xv}{{\bf x}}

\newcommand{\zv}{{\bf z}}
\newcommand{\zerov}{{\bf 0}}

\newcommand{\Am}{{\bf A}}

\newcommand{\Hm}{{\bf H}}
\newcommand{\Id}{{\bf I}}

\newcommand{\Vm}{{\bf V}}

\newcommand{\Ac}{{\cal A}}

\newcommand{\Cc}{{\cal C}}
\newcommand{\Dc}{{\cal D}}

\newcommand{\Lc}{{\cal L}}

\newcommand{\Nc}{{\cal N}}

\newcommand{\Pc}{{\cal P}}

\newcommand{\Rc}{{\cal R}}
\newcommand{\Sc}{{\cal S}}

\newcommand{\Deltam}{\hbox{\boldmath$\Delta$}}

\newcommand{\trace}{{\hbox{tr}}}

\renewcommand{\arg}{{\hbox{arg}}}

\newcommand{\herm}{{\sf H}}

\newcommand{\eq}[1]{(\ref{#1})}
\newcommand{\her}{{\sf H}}

\def\BibTeX{{\rm B\kern-.05em{\sc i\kern-.025em b}\kern-.08em
    T\kern-.1667em\lower.7ex\hbox{E}\kern-.125emX}}

\def\LargeImgWidth{16cm}

\begin{document}

\title{Multiuser MIMO Achievable Rates with Downlink Training and Channel State Feedback}

\author{\authorblockN{Giuseppe Caire,}
\authorblockA{University of Southern California\\
Los Angeles CA, 90089 USA\\
} \and
\authorblockN{Nihar Jindal,}
\authorblockA{University of Minnesota \\
Minneapolis MN, 55455 USA\\
} \and
\authorblockN{Mari Kobayashi,}
\authorblockA{SUPELEC \\
Gif-sur-Yvette, France\\
} \and
\authorblockN{Niranjay Ravindran,}
\authorblockA{University of Minnesota \\
Minneapolis MN, 55455 USA\\
} }

\maketitle

\begin{abstract}
We consider a MIMO fading broadcast channel and compute achievable
ergodic rates when channel state information is acquired at the
receivers via downlink training and it is provided to the transmitter
by channel state feedback.
Unquantized (analog)  and quantized (digital) channel state feedback
schemes are analyzed and compared under various assumptions.
Digital feedback is shown to be potentially superior when the feedback {\em channel uses}
per channel state coefficient is larger than 1.
Also, we show that by proper design of the digital
feedback link, errors in the feedback have a minor effect even if simple uncoded modulation is used on
the feedback channel. We discuss first the case of an unfaded AWGN feedback channel with
orthogonal access and then the case of fading MIMO multi-access (MIMO-MAC).
We show that by exploiting the MIMO-MAC nature of the uplink channel,
a much better scaling of the feedback channel resource with the number of base station antennas
can be achieved. Finally, for the case of delayed feedback,
we show that in the realistic case where the fading process has (normalized) maximum Doppler frequency
shift $0 \leq F < 1/2$,  a fraction $1 - 2F$ of the optimal
multiplexing gain is achievable. The general conclusion of this work
is that {\em very significant} downlink throughput is achievable
with {\em simple and efficient} channel state feedback, provided
that the feedback link is properly designed.
\end{abstract}


\section{Introduction} \label{intro}

In the downlink of a cellular-like system, a base station equipped
with multiple antennas communicates with a number of terminals, each
possibly equipped with multiple receive antennas. If a traditional
orthogonalization technique such as TDMA is used, the base station
transmits to a single receiver on each time-frequency resource and
thus is limited to point-to-point MIMO techniques
\cite{Foschini_Gans, telatar1999cma}. Alternatively, the base
station can use multi-user MIMO
to \emph{simultaneously} transmit to multiple
receivers on the same time-frequency resource. Under the assumption of perfect
channel state information at the transmitter (CSIT) and at the
receivers (CSIR), a combination of single-user Gaussian codes,
linear beamforming and ``Dirty-Paper Coding'' (DPC) \cite{DPC} is
known to achieve the capacity of the MIMO downlink channel
\cite{caire2003atm,vishwanath2003dar,
viswanath2003scv,yu2004scg,weingarten2006crg}.
When the number of base station antennas is larger than the number of
antennas at each terminal, the capacity of the MIMO downlink channel is
significantly larger than the rates achievable with point-to-point
MIMO techniques \cite{caire2003atm, Jindal_Goldsmith_DPC,single-to-multiuser}.

Given the widespread applicability of the MIMO downlink channel
model (e.g., to cellular, WiFi, and DSL), it is of great interest to
design systems that can operate near the capacity limit. Although
realizing the optimal DPC coding strategy still remains a formidable
challenge (see for example \cite{Erez2004,benattan,xiong}), it has
been shown that linear beamforming without DPC performs quite close
to capacity when combined with user selection, again under the
simplifying assumption of perfect channel state information (see for
example \cite{YooGoldsmith06,DS05}).

In real systems, however, channel state information is not \textit{a
priori} provided and must be acquired, e.g., through training.
Acquiring the channel state is a challenging and resource-consuming
task in time-varying systems, and the obtained information is
inevitably imperfect.   It is therefore critical to understand what
rates are achievable under realistic channel state information
assumptions, and in particular to understand the sensitivity of
achievable rates to such imperfections. To emphasize the importance
of channel state information, note that in the extreme case of no
CSIT at the BS and identical fading statistics (and perfect CSIR) at
all terminals, the multi-user MIMO benefit is completely lost and
point-to-point MIMO becomes optimal \cite{caire2003atm}.

\subsection{Contributions of this work}

The focus of this paper is a rigorous information theoretic
characterization of the {\em ergodic} achievable rates of a fading
multiuser MIMO downlink channel in which the UTs and the BS obtain
imperfect CSIR/CSIT via downlink training and channel state
feedback.\footnote{Since this work considers feedback schemes where
the role of transmitter and receiver are reversed, we avoid using
``transmitter'' and ``receiver''  and prefer the use of BS and UT
instead, in order to avoid ambiguity.} Converse results on the
capacity region of the MIMO broadcast channel with imperfect channel
knowledge are essentially open (see for example
\cite{Lapidoth-Shamai} and \cite{ShamaiCaireJindal07} for some
partial results). Here, we focus on the achievable rates of a
specific signaling strategy, zero-forcing (ZF) linear beamforming.
Consistently with contemporary wireless system technology, we assume
that each UT estimates its own channel during a downlink training
phase and then feeds back its estimate over the reverse uplink
channel to the BS.  The BS designs beamforming vectors on the basis
of the received channel feedback, after which an additional round of
downlink training is performed (essentially to inform the UTs of the
selected beamformers). Our results tightly bound the rate that is
achievable after this process in terms of the resources (i.e.,
channel symbols) used for training and feedback and the channel
feedback technique.

The analysis of this paper inscribes itself in the line of works dealing with
``training capacity'' \cite{hassibi-hochwald-03it} of block-fading channels.
Several previous and concurrent works have treated training and
channel feedback for point-to-pont MIMO systems (see for example
\cite{narula,love2003gbm,mukkavilli2003bfr,Jafar1,Jafar2,Love1,Love2})
and, more recently, for  MIMO broadcast channels (see for
example
\cite{Jindal,marzetta2006mtr,FastCSI,ding2007mab,huang:sdma, dana-sharif-hassibi06,
mari-jsac}). However, this paper presents a number of novelties relative to prior/concurrent work:
\begin{itemize}
\item
Rather than assuming perfect CSIR at the UT's, we consider
the realistic scenario where the UTs have imperfect CSIR obtained via
downlink training.  Because the imperfect CSIR is the basis for
the channel feedback from the UTs, this degrades the quality
of the CSIT provided to the BS in a non-negligible manner.

\item
Instead of idealizing the feedback channel as a fixed-rate, error-free bit-pipe,
we explicitly consider transmission from
each UT to the BS over the noisy feedback channel.  This reveals the fundamental joint source-channel coding
nature of channel feedback.  In addition, this allows us to meaningfully measure the
uplink resources dedicated to channel feedback and also allows for a
comparison between analog (unquantized) and digital (quantized)
feedback. We begin by modeling the feedback channel as an
AWGN channel (orthogonal across UTs), and later generalize to a
multiple-antenna uplink channel that is shared by
the UTs. In this way, we precisely quantify the fundamental advantage
of using the multiple BS antennas for efficient channel state feedback.

\item
A fundamental property of the system is that UTs are unaware of the chosen beamforming vectors, because
the beamformers depend on all channels whereas each UT only has an estimate of its own channel.
Several previous works (e.g., \cite{Jindal,yoo2007jsac,swann2006isit})
have resolved this uncertainty by making the unstated assumption that each UT
has perfect knowledge of the post-beamforming SINR.
In contrast, we make no such assumption and rigorously show that this ambiguity can be resolved
by an additional round of (dedicated) training.

\item Most prior work has used a worst-case uncorrelated noise argument \cite{Medard}\cite{lapidoth2002fcp}\cite{hassibi-hochwald-03it} to show that imperfect CSI, at worse, leads to the introduction of additional Gaussian noise and thus the achievable rate is lower bounded by the mutual information with ideal channel state information
and reduced SNR. In our case, however, this same argument yields a largely uncomputable quantity and a
further step must be taken that yields a tractable lower bound in terms of the rate difference between the ideal and actual cases, rather than in terms of a SNR penalty.

\item
We consider delayed feedback and quantify in a simple and appealing form
the loss of degrees of freedom (pre-log factor in the achievable rate) in terms of the fading channel
Doppler bandwidth, which is ultimately related to UT velocity.

\end{itemize}

The analysis presented in this paper is relevant from at least two related but different
viewpoints. On one hand, it provides accurate bounds on the achievable ergodic rates of the
linear zero-forcing beamforming scheme with realistic channel estimation and feedback.
These bounds are useful {\em at any operating SNR} (not necessarily large), and in subsequent
work have been used to optimize the system resources allocated for training and feedback
\cite{kobayashi2008mta}\cite{kobayashi2009itw}.
On the other hand, it yields  sufficient conditions on the training and feedback such that the system
achieves the same {\em multiplexing gain} (also referred to as
``pre-log factor'', or ``degrees of freedom'') of the optimal DPC-based scheme under
perfect CSIR/CSIT. Perhaps the most striking fact about this second aspect is that
the full multiplexing gain of the ideal MIMO broadcast channel
can be achieved with simple pilot-based channel estimation and
feedback schemes that consume a relatively small fraction of the system capacity.
Indeed, a fundamental property of the MIMO broadcast
channel is that the quality of the CSIT must increase with signal-to-noise ratio (SNR),
regardless of what coding strategy is used, in order for the full
multiplexing gain to be achievable \cite{Lapidoth-Shamai,
ShamaiCaireJindal07}.
Under the reasonable assumption that the uplink channel quality is in some sense
proportional to the downlink channel, our work shows that this requirement can be
met using a fixed number of downlink and uplink channel symbols
(i.e., system resources used for training and feedback need not increase with SNR).

When there is a significant delay in the feedback loop, the simple scheme analyzed in this paper does
not attain full multiplexing gain. However, for fading processes with normalized Doppler bandwidth $F$ strictly
less than $1/2$, we show the achievability of a multiplexing gain
equal to $M(1 - 2F)$, where $M$ is the number of BS antennas.
This result follows from a fundamental property of the
noisy prediction error of the channel process and is closed related to Lapidoth's
high-SNR capacity of single-user fading channels without the perfect CSIR assumption \cite{lapidoth2005acs}.

The paper is organized as follows. Section \ref{model.sect}
introduces the system model, describes linear beamforming, and
defines the baseline estimation,  feedback, and beamforming
strategy. Section \ref{rategap.sect} develops bounds on the ergodic
rates achievable by the baseline scheme. In Section \ref{awgn.sect}
we consider an AWGN feedback channel and particularize the rate
bounds to analog and digital feedback (incorporating the effect of
decoding errors for digital feedback), and compare the different
feedback options. Section \ref{sec:fading} generalizes the results
to the setting where the feedback link is a fading MIMO multiple
access channel (MAC). Section \ref{feedback-delay.sect} considers
time-correlated fading and the effect of delay in the feedback link.
Some concluding remarks are provided in Section
\ref{conclusions.sect}.

\section{System Model} \label{model.sect}

We consider a multi-input multi-output (MIMO) Gaussian broadcast channel modeling
the downlink of a system where a Base Station (BS) has $M$ antennas and $K$ User Terminals (UTs) have one antenna each.
A channel use of such channel is described by
\begin{equation} \label{model}
y_k = \hv_k^\her \xv + z_k, \;\; k = 1,\ldots,K
\end{equation}
where $y_k$ is the channel output at UT $k$,
$z_k \sim \Cc\Nc(0,N_0)$ is the corresponding
Additive White Gaussian Noise (AWGN),
$\hv_k \in \CC^M$ is the vector of channel coefficients from the $k$-th UT antenna
to the BS antenna array (the superscript $\her$ refers to the Hermitian, or conjugate transpose)
and $\xv$ is the vector of channel input symbols transmitted by the BS.
The channel input is subject to the average power constraint $\EE[|\xv|^2] \leq P$.

We assume that the channel {\em state}, given by the collection of
all channel vectors $\Hm = [\hv_1,\ldots,\hv_K] \in \CC^{M \times
K}$, varies in time according to a block-fading model \cite{biglieri1998fci}, where $\Hm$
is constant over each {\em frame} of length $T$ channel uses,
and evolves from frame to frame according to an ergodic stationary spatially
white jointly Gaussian process, where the entries of $\Hm$ are Gaussian i.i.d. with elements
$\sim \Cc\Nc(0,1)$. Our bounds on the ergodic achievable rate do not directly depend on the frame
size $T$; rather, these bounds depend only on whether the training, feedback, and data phases
all occur within a frame or in different frames.
In Sections \ref{awgn.sect} - \ref{sec:fading} we consider the simplified scenario where the three
phases all occur within a single frame (i.e., the channel is constant across the phases) and
fading is independent across blocks, but we remove these simplifications in Section \ref{feedback-delay.sect}.
It should also be noticed that the rate lower bounds given in the following
should be multiplied by the factor $(1 - \Delta/T)$, where $\Delta$ denotes the total number of channel uses per frame
dedicated to training and feedback. This factor is neglected in this paper since it is common to all rate bounds and
since $\Delta \ll T$ in a typical slowly-fading system scenario. However,  in the general case where $\Delta$ is not necessarily
small with respect to $T$,  the amount of training and feedback should be optimized by taking this multiplicative factor
into account. Based on the bounds developed in the present paper, this system optimization is carried out in
the follow-up works  \cite{kobayashi2008mta,kobayashi2009itw}.

\subsection{Linear beamforming}

Because of simplicity and robustness to non-perfect CSIT, simple
{\em linear precoding} schemes with standard Gaussian coding have
been extensively considered:  the transmit signal is formed as $\xv = \Vm \uv$,
such that $\Vm \in \CC^{M \times K}$ is a {\em linear beamforming}
matrix and $\uv \in \CC^K$ contains the symbols from $K$ independently generated Gaussian codewords.
In particular, for $K \leq M$ Zero-Forcing  beamforming chooses the $k$-th column $\vv_k$ of $\Vm$
to be a unit vector orthogonal to the subspace $\Sc_k = {\rm span}\{\hv_j : j \neq k\}$.

We focus on the achievable {\em ergodic} rates under ZF linear beamforming and
Gaussian coding. In this case, the achievable rate-sum is given by
\begin{equation} \label{RsumZF}
\max_{\sum_k \EEsmall[\Pc_k(\Hm)] \leq P} \;
\sum_{k=1}^K \EE\left [ \log \left (1 + \frac{|\hv_k^\her\vv_k|^2
\Pc_k(\Hm)}{N_0} \right ) \right ].
\end{equation}
where the optimal power allocation is obtained by waterfilling over the set of channel gains
$\{|\hv_k\vv_k|^2:k=1,\ldots,K\}$. Performance can further be improved by using a {\em user scheduling} algorithm
to select in each frame an {\em active} user subset not larger than $M$ (if $K > M$, such selection must be
performed if ZF is used). Schemes for user scheduling have been extensively discussed, for example
in \cite{yoo2006omb,mari-jsac,DS05,AmirCISS}.

We focus, however, on the case $K=M$ with uniform power allocation (across users and frames: $\Pc_k(\Hm) = \frac{P}{M}$)
and without user selection, in which case the per-user ergodic rate is
\begin{equation} \label{RkZF-uniform}
R_k^{\rm ZF}(P)  = \EE\left [ \log \left (1 + \frac{|\hv_k^\herm \vv_k|^2
P}{N_0M} \right ) \right ] .
\end{equation}
Because $\hv_k$ is spatially white and $\vv_k$ is selected independent of
$\hv_k$ (by the ZF procedure), it follows that $\hv_k^\herm \vv_k$ is $\sim \Cc\Nc(0,1)$.
As a result, $R_k^{\rm ZF}$ is the ergodic capacity of a point-to-point channel in Rayleigh fading with
average SNR $\frac{P}{N_0 M}$, and thus can be written in closed form as \cite{alouini1999crf}
$R_k^{\rm ZF} = \exp \left (\frac{N_0 M}{P} \right ) {\rm E}_i\left (1, \frac{N_0 M}{P} \right )$
where ${\rm E_i}(n,x) = \int_1^\infty \frac{e^{-xt}}{t^n} dt, \; x >
0$ \cite{abramowitz1965hmf}. In the remainder of the paper $R_k^{\rm ZF}$ serves as a benchmark aginst which
we compare the achievable rates with imperfect CSI.

This restriction is dictated by a few reasons. On one hand, the case
$K = M$ without selection makes closed-form analysis (in the
presence of imperfect CSI) possible.  In addition, the maximum
multiplexing gain is $M$ for all $K \geq M$ and hence the case $K =
M$ suffices to capture the fundamental aspects of the problem
(particularly at high SNR). Finally, recent results
\cite{yoo2007jsac,RavindranJindal07} show that the dependence on CSI
quality is roughly the same even when user selection is performed.

\subsection{Channel state estimation and feedback} \label{fbmodel.sect}

We assume that each UT estimates its channel vector from {\em
downlink training symbols} and then feeds this information back to
the BS. This scenario, referred to as ``closed-loop'' CSIT
estimation, is relevant for Frequency-Division Duplexed (FDD)
systems. Our baseline system is depicted in Fig.~\ref{fig-fbmodel}
and consists of the following phases:

\begin{figure}[t]
\begin{center}
\includegraphics[width=\LargeImgWidth]{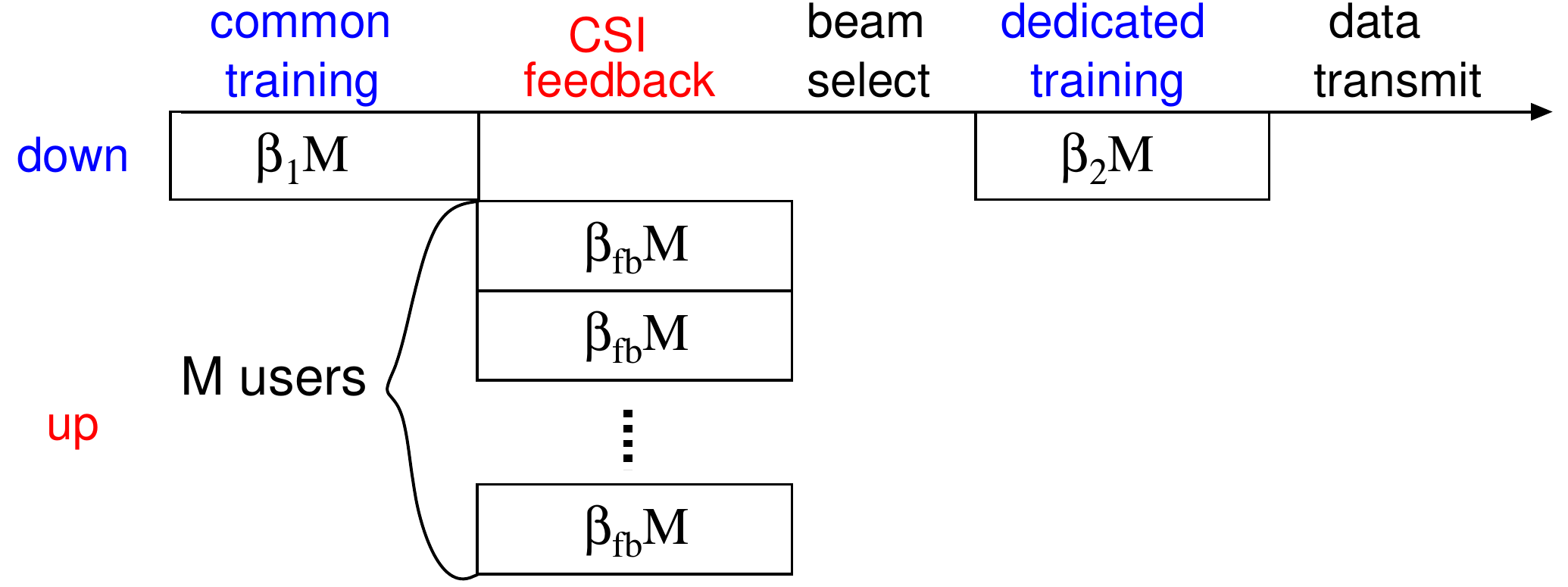}
\vspace{0.5cm} \caption{Channel estimation and feedback
model}\label{fig-fbmodel}
\end{center}
\vspace{-20pt}
\end{figure}

\begin{enumerate}
\item {\bf Common Training}:
~The BS transmits $\beta_1 M$ shared pilots ($\beta_1 \geq 1$ symbols per antenna) on the
downlink\footnote{If $\beta_1$ is an integer, pilot symbols can be orthogonal in time, i.e.,
$\beta_1$ pilots are successively transmitted from each of the $M$ BS antennas for a total of $\beta _1 M$ channel uses.
More generally, it is sufficient for $\beta_1 M$ to be an integer and to use
a unitary $M \times \beta_1 M$ spreading matrix as described in \cite{FastCSI};
in either case the effective received SNR is $\beta_1 \frac{P}{N_0}$.}.
Each UT $k$ estimates its channel from the observation
\begin{equation} \label{training-phase-1-rx}
\sv_k = \sqrt{\beta_1 P}\ \hv_k + \zv_k
\end{equation}
corresponding to the common training (downlink) channel output,
where $\zv_k \sim \Cc\Nc(0,N_0 \Id)$.  The MMSE estimate $\widetilde{\hv}_k$ of $\hv_k$ given the observation $\sv_k$
is given by \cite{poor-book}:
\begin{equation}
\widetilde{\hv}_k = \EE[ \hv_k \sv_k^\herm] \EE[\sv_k\sv_k^\herm]^{-1} \sv_k = \frac{\sqrt{\beta_1 P}}{N_0 + \beta_1 P} \sv_k
\label{train-phase-1-mmse}
\end{equation}
The channel $\hv_k$ can be written in terms of the estimate $\widetilde{\hv}_k$ and estimation noise $\nv_k$ as:
\begin{equation} \label{train-phase-1}
\hv_k = \widetilde{\hv}_k + \nv_k,
\end{equation}
where $\nv_k$ is independent of the estimate and is Gaussian with covariance $\sigma_1^2 \Id$ with
\begin{eqnarray} \label{estimation-error-common-phase}
\sigma_1^2 & = & \frac{1}{1 + \beta_1 P/N_0}
\end{eqnarray}

\item {\bf Channel State Feedback}:
~ Each UT feeds back its channel estimate $\widetilde{\hv}_k$ to the
BS immediately after completion of the common training phase. We use
$\widehat{\Hm} = [\widehat{\hv}_1,\ldots,\widehat{\hv}_K] \in \CC^{M
\times K}$ to denote the (imperfect) CSIT available at the BS; the
feedback is thus a mapping, possibly probabilistic, from
$\widetilde{\hv}_k$ to $\widehat{\hv}_k$. For now we leave the
feedback scheme unspecified to allow development of general
achievability bounds in Section \ref{rategap.sect}, and
particularize to specific feedback schemes from Section
\ref{awgn.sect} onwards.

In Section \ref{awgn.sect} we consider the simplified setting where
the feedback channel is an unfaded AWGN channel SNR $\frac{P}{N_0}$, orthogonal
across UTs, but in Section \ref{sec:fading} we consider the more realistic
setting where the uplink channel is a MIMO-MAC with fading.
Furthermore, the baseline model of Fig.~\ref{fig-fbmodel} assumes no delay
in the feedback, i.e., the channel is constant across the training, feedback, and data phases.
In Section \ref{feedback-delay.sect} we remove this assumption and consider the
case where feedback has delay and the channel state changes from
frame to frame according to a time-correlation model.

We assume each UT transmits its feedback over $\beta_{\rm fb} M$ feedback channel symbols.

\item {\bf Beamformer Selection}:
~ The BS selects the beamforming vectors by treating the estimated
CSIT $\widehat{\Hm}$ as if it was the true channel (we refer to this approach as ``naive'' ZF beamforming).
Following the ZF recipe,
$\widehat{\vv}_k$ is a unit vector orthogonal to the subspace $\Sc_k
= {\rm span}\{\widehat{\hv}_j : j \neq k\}$. We use the notation
$\widehat{\Vm} = [\widehat{\vv}_1,\ldots,\widehat{\vv}_K]$.
Since $K=M$ and the BS channel estimates $\widehat{\hv}_1, \ldots,
\widehat{\hv}_M$ are independent, the subspace $\Sc_k$ is $M-1$
dimensional (with probability one) and is independent of
$\widehat{\hv}_k$.  The beamforming vector $\widehat{\vv}_k$ is chosen in the
one-dimensional nullspace of $\Sc_k$; as a result $\widehat{\vv}_k$ is independent of the channel
estimate $\widehat{\hv}_k$ and of the true channel vector $\hv_k$.

\item {\bf Dedicated Training}:
~ Once the the BS has computed the beamforming vectors
$\widehat{\Vm}$, coherent detection of data at each UT is enabled by an additional round of downlink training
transmitted along each beamforming vector. This additional round of training is required
because the beamforming vectors $\{\widehat{\vv}_k\}$ are functions
of the channel state information $\{\widehat{\hv}_1,\ldots,\widehat{\hv}_K\}$ at the BS, while UT $k$
knows only $\widetilde{\hv}_k$ or, at best, $\widehat{\hv}_k$ (if error-free digital feedback is used).
Therefore, the coupling coefficients between the
beamforming vectors and the UT channel vector are unknown.

Let the set of the coefficients affecting the signal received by UT $k$ be denoted by
\[ \Ac_k \triangleq \{a_{k,j}: j = 1,\ldots,M\} \]
where $a_{k,j} = \hv_k^\herm \widehat{\vv}_j$ is the coupling
coefficient between the $k$-th channel and the $j$-th beamforming vector.
The received signal at the $k$-th UT is given by
\begin{eqnarray} \label{receiver-k}
y_k  =  \hv^\herm_k \widehat{\Vm} \uv + z_k
&=& a_{k,k} u_k + \sum_{j \neq k} a_{k,j} u_j + z_k \nonumber \\
& = & a_{k,k} u_k + I_k + z_k
\end{eqnarray}
where the interference at UT $k$ is denoted as
\begin{equation}
I_k = \sum_{j\neq k} a_{k,j} u_j
\end{equation}
and $a_{k,k}$ is the
{\em useful signal} coefficient. The dedicated training is intended to allow the estimation of
the coefficients in $\Ac_k$ at each UT $k$.
This is accomplished by transmitting $\beta_2$ orthogonal training symbols along each of the beamforming
vectors on the downlink, thus requiring a total of $\beta_2 M$ downlink channel uses.\footnote{If $\beta_2 M$ is an integer but $\beta_2$ is not,
the unitary spreading approach used for common training can also be used here.}
The relevant observation model for the estimation of $\Ac_k$ is given by
\begin{equation} \label{dedicated-training-obs}
r_{k,j} = \sqrt{\beta_2 P}\ a_{k,j} + z_{k,j} , \;\; j = 1,\ldots,M
\end{equation}
We denote the full set of observations available to UT $k$ as:
\[ \Rc_k \triangleq \{r_{k,j} : j = 1,\ldots,M \}. \]
In particular, we shall consider explicitly the case where UT $k$ estimates its useful signal coefficient
using linear MMSE estimation based on $r_{k,k}$, i.e.,
\begin{equation} \label{akhat}
\widehat{a}_{k,k} = \frac{\sqrt{\beta_2P}}{N_0 + \beta_2P} r_{k,k}.
\end{equation}
Because $\widehat{\vv}_k$ is a unit vector independent of $\hv_k$,
the useful signal coefficient  $a_{k,k} = \hv_k^\herm
\widehat{\vv}_k$ is complex Gaussian with unit variance.  As a
result we have the representation
\begin{equation} \label{train-phase-2}
a_{k,k} = \widehat{a}_{k,k} + f_k
\end{equation}
where $f_k$ and $\widehat{a}_{k,k}$ are independent and Gaussian with variance
$\sigma_2^2$ and $1 - \sigma_2^2$, respectively, with
\begin{equation} \label{estimation-error-dedicated-phase}
\sigma_2^2 = \frac{1}{1 + \beta_2 P/N_0}.
\end{equation}

\item {\bf Data Transmission}:
After the dedicated downlink training phase, the BS sends the coded data symbols
$u_1,\ldots,u_K$ for the rest of the frame duration.
The effective channel output for this phase is therefore given by the sequence of corresponding channel output symbols $y_k$ given by \eq{receiver-k},
and by the observation of the dedicated training phase $\Rc_k$ given by
(\ref{dedicated-training-obs}).

When considering the {\em ergodic} rates achievable by the proposed scheme,
we implicitly assume that coding is performed over a long sequence of frames,
each frame comprising a common training phase, channel state feedback phase,
dedicated training phase and data transmission.
\end{enumerate}

\vspace{11pt} \noindent We conclude this section with a few remarks.
First, we would like to observe that
two phases of training, a common ``pilot channel'' and dedicated per-user training symbols is
common practice in some wireless cellular systems, as for example in
the downlink of the 3rd generation Wideband CDMA standard
\cite{prasad2000tgm} and in the MIMO component of future
4th generation systems \cite{pilot-patent}.
Second, we note that an alternative to FDD is Time-Division
Duplexing (TDD), where uplink and downlink share in time-division
the same frequency band. In this case, provided that the coherence
time is significantly larger than the concatenation of an uplink and
downlink slot and hardware calibration, the downlink channel can be
learned by the BS from uplink training symbols \cite{FastCSI,
jose:csi}. Although we focus on FDD systems, in Remark
\ref{remark-tdd} we note the straightforward extension of our
results to TDD systems.

\section{Achievable Rate Bounds} \label{rategap.sect}

We assume that the user codes are independently generated
according to an i.i.d. Gaussian distribution, i.e., the input symbols are $u_k \sim \Cc\Nc(0,P/M)$.
The remainder of this section is dedicated to deriving upper and lower bounds on the mutual
information achieved by such Gaussian inputs, indicated by $R_k \triangleq I(u_k; y_k, \Rc_k)$.

\subsection{Lower Bounds}

The following lower bound is obtained by using techniques similar to those
in \cite{Medard,hassibi-hochwald-03it,lapidoth2002fcp}.

\begin{thm} \label{mutual-info-bound.theorem}
The achievable rate for ZF beamforming with Gaussian inputs and CSI training and
feedback as described in Section \ref{fbmodel.sect} can be bounded from below by:
\begin{equation} \label{mutual-info-bound}
R_k \geq \EE \left[ \log\left( 1 + \frac{|\widehat{a}_{k,k}|^2 P/(N_0 M)}
{1 + \sigma_2^2 P/(N_0 M)  + \EE\LSB |I_k|^2| \widehat{a}_{k,k} \RSB /N_0 }\right) \right]
\end{equation}
\end{thm}
{\bf Proof:} See Appendix \ref{proof_thm1}. \hfill $\square$

The conditional interference second moment $\EE\LSB |I_k|^2|\widehat{a}_{k,k}\RSB$
in (\ref{mutual-info-bound}) may be difficult to compute even by Monte Carlo simulation,
due to the complicated dependency of $I_k$ on $\widehat{a}_{k,k}$ (this dependence is unknown even
if the dedicated training is perfect, i.e., $\widehat{a}_{k,k} = a_{k,k}$).
However, we will not need to compute this explicitly, as is seen in our next results.

A very useful measure is the difference between $R_k$ and $R_k^{\rm ZF}$,
the achievable rate with ZF beamforming and ideal CSI defined in (\ref{RkZF-uniform}).
The rate gap is defined as follows
\begin{equation} \label{delta-r-def}
\Delta R \define R_k^{\rm ZF} - R_k,
\end{equation}
and is upper bounded in the following theorem.
\begin{thm} \label{rate-gap-bound.theorem}
The rate gap incurred by ZF beamforming with training and feedback as described in
Section \ref{fbmodel.sect} with respect to ideal ZF with equal power allocation is upperbounded by:
\begin{equation} \label{rate-gap-bound}
\Delta R  \leq \log \left ( 1 + \sigma_2^2\ \frac{P}{N_0M} + \frac{\EE[|I_k|^2]}{N_0} \right )
\end{equation}
\end{thm}

{\bf Proof:}  See Appendix \ref{proof_thm2}. \hfill $\square$

For clarity of notation, we denote the RHS of the above, referred to as the rate gap upper bound, as $\overline{\Delta R}$:
\begin{eqnarray}
\overline{\Delta R}  &\triangleq& \log \left ( 1 + \sigma_2^2\ \frac{P}{N_0M} + \frac{\EE \LSB |I_k|^2 \RSB}{N_0} \right ) \\
&=& \log \left ( 1 + \frac{P}{N_0 M} \left( \sigma_2^2\  + \sum_{j \neq k} \EE \LSB |\hv_k^\herm \widehat{\vv}_j|^2 \RSB \right)
\right ) \label{eq-rategap_simple}
\end{eqnarray}
where the latter follows from a simple calculation of $\EE [|I_k|^2]$.
The term $\sigma_2^2$ depends only on dedicated training;
on the other hand, $\EE \LSB |\hv_k^\herm \widehat{\vv}_j|^2 \RSB$ is determined by the mismatch
between $\hv_k$ and the BS estimate $\widehat{\hv}_k$
(because $\widehat{\vv}_j$ is chosen orthogonal to $\widehat{\hv}_k$ rather than $\hv_k$)
and therefore depends on the common training and feedback phases.

An obvious result of the rate gap upper bound is the following lower bound to $R_k$:
\begin{cor} \label{cor-lower}
The achievable rate for ZF beamforming with Gaussian inputs and CSIT training and
feedback as described in Section \ref{fbmodel.sect} can be bounded from below by:
\begin{equation}
R_k \geq
R_k^{\rm ZF} - \overline{\Delta R}
\end{equation}
\end{cor}
\vspace{5pt}
Because only the estimate of $a_{k,k}$ is used in the derivation,
Corollary \ref{cor-lower} is also a lower bound to $I(u_k; y_k, r_{k,k})$.

\subsection{Upper Bounds}

A useful upper bound to $R_k$ is reached by providing each UT $k$
with exact knowledge of the interference
coefficients $\Ac_k$. Thus, this is referred to as the ``genie-aided upper-bound''.
\begin{thm} \label{genie-mutual-info-bound.theorem}
The achievable rate for ZF beamforming with Gaussian inputs and CSI training and
feedback is upper bounded by the rate achievable when, after the beamforming matrix
$\widehat{\Vm}$ is chosen, a genie provides the $k$-th UT with perfect
knowledge of the coefficients $\Ac_k = \{a_{k,j} = \hv_k^\herm \widehat{\vv}_j : j = 1,\ldots,M\}$:
\begin{equation} \label{genie-mutual-info-bound}
R_k \leq \EE \left[ \log\left( 1 + \frac{|a_{k,k}|^2 P/(N_0 M)} {1 + \sum_{j \neq k} |a_{k,j}|^2 P/(N_0 M)} \right) \right].
\end{equation}
\end{thm}
\vskip 12pt
{\bf Proof:}
Since $\Rc_k$ is a noisy version of $\Ac_k$, the data-processing inequality yields
\begin{equation}
R_k = I(u_k; y_k, \Rc_k) \leq I(u_k;y_k, \Ac_k)
\end{equation}
Because $y_k$ conditioned on $\Ac_k$ is
complex Gaussian with variance $N_0 + \sum_{j=1}^M |a_{k,j}|^2 P/M$
while $y_k$ conditioned on $(\Ac_k, u_k)$ is
complex Gaussian with variance $N_0 + \sum_{j \ne k} |a_{k,j}|^2 P/M$, we immediately obtain
(\ref{genie-mutual-info-bound}).
\QED

The practical relevance of Theorem \ref{genie-mutual-info-bound.theorem} is two-fold: on one hand,
(\ref{genie-mutual-info-bound}) is easy to evaluate by Monte Carlo simulation.\footnote{It is usually difficult if not impossible to obtain in closed form the joint distribution of the coefficients  $\Ac_k$.}
On the other hand, this bound can be approached for large $\beta_2$,  since in this case each UT can accurately
estimate all interference coupling coefficients and not only the useful  signal coefficient.

\section{Channel state feedback over an AWGN Channel} \label{awgn.sect}

In this section we quantify the rate gap upper bound for different feedback strategies under the assumption that the feedback channel is an
unfaded AWGN channel with the same SNR as the downlink, i.e., $P/N_0$, and that the UTs access the channel
orthogonally.  Each UT uses $\beta_{\rm fb} M$ feedback channel symbols, and therefore
the total number of feedback channel uses is $\beta_{\rm fb} M^2$.

\subsection{Analog feedback} \label{analog.sect}

Analog feedback refers to transmission (on the feedback link) of the estimated downlink channel coefficients
by each UT using unquantized quadrature-amplitude modulation
\cite{FastCSI,mari-jsac,motorola-guys,Mandayam}.
More specifically, each UT transmits on the feedback channel
a scaled version of its common downlink training observation $\sv_k$
defined in (\ref{training-phase-1-rx}). The resulting feedback channel output (BS observation)
relative to UT $k$ is given by:
\begin{eqnarray}
\gv_k & = &\frac{\sqrt{\beta_{\rm fb} P}}{\sqrt{\beta_1P + N_0}}\ {\sv_k} + \widetilde{\wv}_k  \label{analog-feedback-1}\\
& = &\frac{\sqrt{\beta_{\rm fb}\beta_1} P}{\sqrt{\beta_1P + N_0}}\ \hv_k + \frac{\sqrt{\beta_{\rm fb} P}}{\sqrt{\beta_1P + N_0}}\ \zv_k + \widetilde{\wv}_k  \label{analog-feedback-2a}\\
& = &\frac{\sqrt{\beta_{\rm fb}\beta_1} P}{\sqrt{\beta_1P + N_0}}\ \hv_k +\wv_k \label{analog-feedback-2}
\end{eqnarray}
where $\widetilde{\wv}_k$ represents the AWGN noise on the uplink
feedback channel (variance $N_0$) and $\zv_k$ is the noise during
the common training phase. The power scaling $\beta_{\rm fb}$
corresponds to the number of channel uses per channel coefficient
(we require $\beta_{\rm fb} \geq 1$ so that each coefficient is transmitted at least once),
assuming that transmission in the feedback channel has per-symbol
power $P$ (averaged over frames) and that the channel state vector
is modulated by a $\beta_{\rm fb} M \times M$ unitary spreading
matrix \cite{FastCSI}. Because $\widetilde{\wv}_k$ and $\zv_k$ are
each complex Gaussian with covariance $N_0 {\bf I}$ and are independent,
$\wv_k$ is complex Gaussian with covariance $\sigma_w^2 {\bf I}$ with:
\begin{equation} \label{noise-analog-feedback}
\sigma_w^2 = N_0 \left ( 1 + \frac{\beta_{\rm fb} P/N_0}{1 + \beta_1P/N_0} \right )
\end{equation}
The BS computes the MMSE estimate of the channel vector $\hv_k$ based on $\gv_k$ as:
\begin{equation} \label{analogfb-mmse}
\widehat{\hv}_k = \frac{\sqrt{\beta_{\rm fb}\beta_1} P}{\sqrt{\beta_1P + N_0}\LB \beta_{\rm fb} P + N_0\RB}\ \gv_k .
\end{equation}
Using \eq{analog-feedback-2}, the channel can be written in terms of the
BS estimate and estimation error $\ev_k$ as:
\begin{equation} \label{analogfb-iid}
\hv_k = \widehat{\hv}_k + \ev_k
\end{equation}
where $\ev_k$ is independent of the estimate and is Gaussian with covariance $\sigma_e^2 {\bf I}$ with:
\begin{eqnarray}
\sigma_e^2 = \frac{\sigma_w^2}{\sigma_w^2 + \frac{\beta_{\rm
fb}\beta_1 P^2}{\beta_1P + N_0}} &=& \frac{1}{1 + \beta_{\rm fb}
\frac{P}{N_0}} + \frac{\beta_{\rm fb} \frac{P}{N_0}} {(1 +\beta_{\rm
fb} \frac{P}{N_0})(1 + \beta_1 \frac{P}{N_0})}
\label{error-analog-feedback}.
\end{eqnarray}

This characterization of $(\hv_k, \widehat{\hv}_k)$ can be used
to derive the rate gap upper bound for analog feedback:
\begin{thm} \label{thm:analog_rate_gap}
If each UT feeds back its channel coefficients in analog fashion over
$\beta_{\rm fb} M$ channel uses of an AWGN uplink channel with SNR $\frac{P}{N_0}$,
the rate gap upper bound is given by (``AF'' standing for Analog Feedback):
\begin{equation} \label{analog_agwn_rate_gap}
\overline{\Delta R}^\textsc{AF} =
\log \left ( 1 + \frac{P}{N_0 M}
\left( \frac{1}{1+\beta_2\frac{P}{N_0}} +  (M-1) \left( \frac{1}{1 + \beta_{\rm fb} \frac{P}{N_0}} +
\frac{\beta_{\rm fb} \frac{P}{N_0}}
{(1 +\beta_{\rm fb} \frac{P}{N_0})(1 + \beta_1 \frac{P}{N_0})}  \right) \right)\RB.
\end{equation}
\end{thm}
\vspace{5pt}
{\bf Proof:} See Appendix \ref{proof_thm4}. \QED

It is straightforward to see that $\overline{\Delta R}^\textsc{AF}$ can be upper bounded as
\begin{eqnarray}\label{AFGapAWGN}
\overline{\Delta R}^\textsc{AF}
&\leq&  \log \LB 1 + \frac{1}{M \beta_2} + \frac{M-1}{M} \LB\frac{1}{\beta_1} + \frac{1}{\beta_{\rm fb}}\RB \RB,
\label{analog-awgn-2}
\end{eqnarray}
Hence, the rate gap is uniformly bounded for all SNRs and therefore
the multiplexing gain is preserved (i.e., $\lim_{P \rightarrow \infty} \frac{R_k}{\log_2 P} =
1$) in spite of the imperfect CSI.

An {\em intuitive} understanding of this rate loss is obtained if one re-examines
the UT received signal in the form used in Theorem \ref{mutual-info-bound.theorem}:
\begin{equation}
y_k = \widehat{a}_{k,k}u_k +
\underbrace{f_k u_k}_{\textrm{Self Noise}}  +
\underbrace{\sum_{j\neq k} (\hv_k^\her \widehat{\vv}_j) u_j}_{\textrm{Interference}}
+ \underbrace{z_k}_{\textrm{Noise}}
\end{equation}
The imperfect channel state information (at the UT and BS)
effectively increases the noise from the thermal noise level $N_0$
to the sum of the thermal noise, self-noise, and interference power,
and the rate gap upper bound $\overline{\Delta R}^\textsc{AF}$ is
precisely the logarithm of the ratio of the effective noise to the
thermal noise power.

\begin{remark}
In many systems, the uplink SNR is smaller than the downlink SNR because
UT's transmit with reduced power.  If the uplink SNR is $\Gamma \frac{P}{N_0}$
rather than $\frac{P}{N_0}$, $\overline{\Delta R}^\textsc{AF}$ is equal to the expression
in Theorem \ref{thm:analog_rate_gap} with $\beta_{\rm fb}$ replaced with
$\Gamma \beta_{\rm fb}$.  This does not change the multiplexing gain, but can have a significant effect on the rate gap.
\hfill $\lozenge$
\end{remark}

\begin{remark} \label{remark-tdd}
It is easy to see that a TDD system with perfectly reciprocal uplink-downlink channels where each
UT transmits $\beta_{TDD}$ pilots (a single pilot trains all $M$ BS antennas)
in an orthogonal manner corresponds exactly to an FDD system with perfect feedback
($\beta_{\rm fb} \rightarrow \infty$) and $\beta_1 = \beta_{TDD}$, because the downlink training in an FDD
system is equivalent to the uplink training in a TDD system.
Therefore, as a byproduct of our analysis, we obtain a result for TDD open loop CSIT estimation:
\begin{eqnarray}
\overline{\Delta R}^\textsc{TDD} & = &
\log \left  [ 1 + \frac{P}{N_0 M} \left (\frac{1}{1 + \beta_2 \frac{P}{N_0}} + \frac{M-1}{1 + \beta_{TDD} \frac{P}{N_0}}
\right ) \right ]  \\
& \leq & \log \LB 1 + \frac{1}{M \beta_2} + \frac{M-1}{M} \frac{1}{\beta_{TDD}}\RB.
\end{eqnarray}
Dedicated training is necessary even in TDD systems because UT's do not know
the channels of other UT's and thus are not aware of the beamforming vectors
used by the BS.  Finally, note that in TDD a total of $M \beta_{TDD}$ uplink
training symbols and $M \beta_2$ downlink (dedicated) training symbols are needed.
\hfill $\lozenge$
\end{remark}

\subsection{Digital feedback} \label{digital.sect}

We now consider ``digital'' feedback, where the estimated channel vector is quantized
at each UT and represented by $B$ bits. The packet of $B$ bits is fed back by each UT to the BS.
We begin by computing the rate gap upper bound in terms of {\em bits}, and later in the section relate this to feedback {\em channel uses}.

Following \cite{mukkavilli2003bfr,love2003gbm,narula,Jindal}, we consider a specific scheme for channel state quantization based
on a quantization codebook $\Cc = \{\pv_1, \dots, \pv_{2^B}\}$ of unit-norm vectors in $\CC^M$.
The quantization $\widehat{\hv}_k$ of the estimated channel vector $\widetilde{\hv}_k$ is found according to the decision rule:
\begin{equation} \label{quantization-rule}
\widehat{\hv}_k = \mathop{\arg \max} \limits_{\pv\ \in\ \Cc}\
|\widetilde{\hv}_k^\herm\pv|^2
\end{equation}
and thus $\widehat{\hv}_k$ is the quantization vector forming the minimum angle with $\widetilde{\hv}_k$.
The corresponding $B$-bits quantization index is fed back to the BS.
Because $\widehat{\hv}_k$ is unit-norm, no channel magnitude information is conveyed.

In \cite{Love1, Jindal} it is shown that for a random ensemble of
quantization codebooks referred to as {\em Random Vector
Quantization} (RVQ), obtained by generating $2^B$ quantization
vectors independently and uniformly distributed on the unit sphere
in $\CC^M$ (see \cite{Jindal} and references therein), the average
(angular) distortion is given by:
\begin{equation} \label{RVQ-bound}
\EE\LSB
\sin^2 \left( \widetilde{\hv}_k,  \widehat{\hv}_k \right) \RSB
= 2^B \beta \left( 2^B, \frac{M}{M-1} \right) \leq 2^{-\frac{B}{M-1}}
\end{equation}
where $\beta(\cdot)$ is the beta function and $\sin^2 \left(
\widetilde{\hv}_k,  \widehat{\hv}_k \right) = 1 -
\frac{|\widetilde{\hv}_k^\herm\widehat{\hv}_k|^2}{\|\widetilde{\hv}_k\|^2}$.
As in \cite{Jindal} we assume each UT uses an independently
generated codebook. For this particular quantization scheme, we can
compute the rate gap upper bound:
\begin{thm} \label{thm:digital_rate_gap}
If each UT quantizes its channel to $B$ bits (using RVQ) and conveys these
bits in an error-free fashion to the BS, the rate gap upper bound is given by (``DF'' standing for Digital Feedback):
\begin{equation} \label{eq-digital_upper1}
\overline{\Delta R}^\textsc{DF}  =
\log \LB 1 + \frac{P}{N_0 M} \left( \frac{1}{1 + \beta_2\frac{P}{N_0}} +
\frac{M}{1+\beta_1\frac{P}{N_0}} \left [ \frac{M-1}{M} +
\frac{\beta_1 P}{N_0} 2^B \beta \left( 2^B, \frac{M}{M-1} \right)
\right] \right) \RB.
\end{equation}
\end{thm}
\vspace{5pt}
{\bf Proof:} See Appendix \ref{proof_thm5}. \QED

Using (\ref{RVQ-bound}), the rate gap upper bound is further upper bounded as:
\begin{eqnarray}
\overline{\Delta R}^\textsc{DF} &\leq& \log \LB 1 +
\frac{1}{M\beta_2} + \frac{M-1}{M} \; \frac{1}{\beta_1} +
\left(\frac{P}{N_0}\right) 2^{-\frac{B}{M-1}} \RB
\label{digital-2a}
\end{eqnarray}
Comparing this to the rate gap in the analog feedback case (\ref{analog-awgn-2}),
we notice that the dependence on $\beta_1$ and $\beta_2$ are precisely the same for both analog and
digital feedback.

The next step is translating the rate gap upper bound so that it is in terms of
feedback symbols rather than bits.  For the time being, we shall make the very unrealistic assumption
that the feedback link can operate error-free at capacity, i.e., it can reliably transmit $\log_2(1 + P/N_0)$ bits per
symbol.\footnote{This assumption is unrealistic in the context of this model
because the feedback channel coding block length is very small and because the need for very fast feedback
(essentially delay-free) prevents grouping blocks of channel coefficients and using
larger coding block length.}

The analog feedback considered before provides a noisy version of the
channel vector norm in addition to its direction. Although this
information is irrelevant for the ZF beamforming considered here, it
might be useful in some user selection algorithms such as those
proposed in \cite{yoo2006omb,mari-jsac,DS05,AmirCISS}. In contrast,
digital feedback based on unit-norm quantization vectors provides no norm information.
Thus, for fair comparison, we assume that $\beta_{\rm fb} M$ feedback
symbols in the analog feedback scheme correspond to $\beta_{\rm fb}
(M-1)$ feedback symbols for the digital feedback scheme; i.e., a
system using digital feedback could use one feedback symbol to
transmit channel norm information.  An alternative justification for this is to notice that the analog feedback
system could be modified to operate in $\beta_{\rm fb}(M-1)$ channel symbols
by transmitting only the $M-1$ \emph{relative} phases and amplitudes
of the channel coefficients, since the absolute norm and phase are
irrelevant to the ZF beamforming considered here.

Under this assumption, the number of feedback bits per mobile is
$B = \beta_{\rm fb} (M-1) \log_2 (1 + P/N_0)$.
Plugging this into (\ref{digital-2a}) gives:
\begin{equation} \label{rate-gap-digital-symbols}
\overline{\Delta R}^\textsc{DF} \leq  \log \LB 1 +
\frac{1}{M\beta_2} + \frac{M-1}{M} \; \frac{1}{\beta_1} +
\frac{\frac{P}{N_0}}{\left (1 + \frac{P}{N_0} \right)^{\beta_{\rm
fb}}}  \RB
\end{equation}
Similar to analog feedback, if $\beta_{\rm fb} \geq 1$ then the rate gap is upper bounded and full multiplexing gain
is preserved.  However, it should be noticed that for $\beta_{\rm fb}$  strictly larger than 1 digital feedback
yields a term $\left(\frac{P}{N_0} \right)^{1-\beta_{\rm fb}}$ that  vanishes as $P/N_0 \rightarrow \infty$.
This should be contrasted with the constant  term $\frac{1}{\beta_{\rm fb}}$ for the case of analog feedback.

\subsection{Effects of feedback errors} \label{errors.sect}

We now remove the optimistic assumption
that the digital feedback channel can operate error-free at capacity.
In general, coding for the CSIT feedback channel should be regarded
as a joint source-channel coding problem, made particularly interesting by the non-standard
distortion measure
and by the fact that a very short block length is required.
A thorough discussion of this subject is out of the scope of the present
paper and is the matter of current investigation  (see for example \cite{hooman-TCOM, raj-ISIT09}).
Here, we restrict ourselves to the detailed analysis of a particularly simple scheme
based on uncoded QAM. Perhaps surprisingly, this scheme is {\em sufficient} to achieve a
vanishing rate gap in the high SNR region, for an appropriate choice of the system parameters.

In the proposed scheme, the UTs perform quantization using RVQ and
transmit the feedback bits using plain uncoded QAM.  The
quantization bits are randomly mapped onto the QAM symbols (i.e., no
intelligent bit-labeling or mapping is used). Therefore, even a
single erroneous feedback bit from UT $k$ makes the BS's CSIT vector
$\widehat{\hv}_k$ essentially useless. Also, no particular error
detection strategy is used and thus the BS computes the beamforming
matrix on the basis of the received feedback, although this may be
in error.

We again let $\beta_{\rm fb} (M-1)$ denote the number of channel uses to transmit the
feedback bits (per UT). Interestingly, even for this very simple scheme there is a non-trivial
tradeoff between quantization distortion and channel errors.
In order to maintain a bounded rate gap, the number of feedback bits must be scaled at least as $(M-1) \log_2\LB1 + \frac{P}{N_0}\RB \approx (M-1) \log_2 \frac{P}{N_0}$.
Therefore, we consider sending $B = \alpha (M-1) \log_2 \frac{P}{N_0}$ bits
for $1 \leq \alpha \leq \beta_{\rm fb}$  in $\beta_{\rm fb} (M-1)$ channel uses, which
corresponds to $\frac{\alpha}{\beta_{\rm fb}} \log_2 \frac{P}{N_0}$ bits per QAM symbol.

The symbol error rate for square QAM with $q$ constellation points
is bounded by  \cite{goldsmith2005wc}:
\begin{eqnarray} \label{qam-error-exact}
P_s = 1 - \left(1 - 2 \left( 1 - \frac{1}{\sqrt{q}} \right) Q \left( \frac{ 3 (P/N_0) }{q-1}\right)
\right)^2
\leq  2 \exp \left (- \frac{3}{2} \frac {P/N_0} {q - 1} \right).
\end{eqnarray}
where $Q(x) = \int_x^\infty \frac{1}{\sqrt{2\pi}} e^{-t^2/2} dt$ is the Gaussian probability tail function.
Using the fact that $q = (P/N_0)^{\frac{\alpha}{\beta_{\rm fb}}}$, we obtain
the upper bound
\begin{eqnarray} \label{qam-ser}
P_s &\leq &
 2 \exp \left (- \frac{3}{2} \left (\frac{P}{N_0} \right )^{1 - \frac{\alpha}{\beta_{\rm fb}}} \right )
\end{eqnarray}
If $\alpha = \beta_{\rm fb}$, which corresponds to signaling at
capacity with uncoded modulation, $P_s$ does not decrease with SNR
and system performance is very poor. However, for $\alpha <
\beta_{\rm fb}$, which corresponds to transmitting at a
fraction of capacity, $P_s \rightarrow 0$ as $\frac{P}{N_0}
\rightarrow \infty$. The error probability of the entire feedback
message (transmitted in $\beta_{\rm fb} (M-1)$ QAM symbols) is given by
\begin{eqnarray} \label{vector-error}
P_{e,{\rm fb}} &=& 1 - (1 - P_s)^{\beta_{\rm fb} (M-1)} \leq \beta_{\rm fb} (M-1) P_s,
\end{eqnarray}
where the inequality follows from the union bound.
Note the tradeoff between distortion and feedback error: $\alpha$ large yields
finer quantization but larger $P_{e,{\rm fb}}$, while $\alpha$ small provides
poorer quantization but smaller $P_{e,{\rm fb}}$.

\begin{thm} \label{thm:digital_rate_gap_errors}
If each UT quantizes its estimated channel using $B = \alpha (M-1) \log P/N_0$ bits (using RVQ),
and transmits on the feedback link using $\beta_{\rm fb} (M-1)$  channel uses with uncoded QAM modulation,
the resulting rate gap can be upperbounded by
\begin{eqnarray} \label{rate-fberror_general}
\overline{\Delta R}^\textsc{DF-Errors}  & \leq & \log \left ( 1 +
\frac{1}{M\beta_2} +  (1 - P_{e,{\rm fb}}) \left( \left(\frac{P}{N_0}
\right)^{1 - \alpha} + \frac{M-1}{M} \ \frac{1}{\beta_1}
\right) + \frac{P}{N_0} P_{e,{\rm fb}} \right ),
\end{eqnarray}
where $P_{e,{\rm fb}}$ is given by (\ref{qam-ser}) and (\ref{vector-error}).
\end{thm}
\vspace{5pt}
{\bf Proof:} See Appendix \ref{proof_thm6}. \QED

If $1 < \alpha < \beta_{\rm fb}$, then the effect of feedback vanishes as $\frac{P}{N_0} \rightarrow \infty$,
somewhat similar to the case of error-free feedback. This is because the feedback error probability decays
exponentially as $(P/N_0)^{1 - \frac{\alpha}{\beta_{\rm fb}}}$,
so that the term  $\frac{P}{N_0} P_{e,{\rm fb}}$ vanishes as $\frac{P}{N_0} \rightarrow \infty$ for
all $\alpha < \beta_{\rm fb}$,  while obviously $(P/N_0)^{1 - \alpha}$ vanishes for all $\alpha > 1$.

A number of simple improvements are possible.
For example, each UT may estimate its interference coefficients
$\{a_{k,j} : j \neq k \}$ from the dedicated training phase, and decide if its feedback
message was correctly received or was received in error by setting a
threshold on the interference power:  if the interference power is $ \approx (M - 1)P$, then it is likely that a feedback
error occurred. If, on the contrary, it is $\approx 2^{-B/(M-1)} P$, then it is likely that the feedback message was
correctly received. Interestingly, for $B = \alpha (M-1) \log_2
\frac{P}{N_0}$ with $\alpha > 1$, detecting feedback error events
becomes easier and easier as $\frac{P}{N_0}$ increases and/or as the
number of antennas $M$ increases. In brief, for a large number of
antennas any terminal whose feedback message was received in error
is completely drowned into interference and should be able to detect this event with high probability.
Assuming that the UTs can perfectly detect their
own feedback error events as described above,
then they can simply discard the frames corresponding to feedback errors.
The resulting achievable rate in this case is lowerbounded by
\begin{eqnarray} \label{genie-fberror-lb}
R_k^{\textsc{DF-Errors-Detect}} & \geq & (1 - P_{e,{\rm fb}}) \left [
R_k^{\rm ZF} - \log \left ( 1 + \frac{1}{M \beta_2} + \frac{M-1}{M} \ \frac{1}{\beta_1} +
\left (\frac{P}{N_0}\right )^{1-\alpha} \right ) \right  ]
\end{eqnarray}
in light of (\ref{rate-gap-digital-symbols}) (after replacing $\alpha$ instead of $\beta_{\rm fb}$)
and of Corollary \ref{cor-lower}.   Note that this rate lies between the achievable rate lower bound obtained
via the rate gap in    \eq{rate-fberror_general}  and the genie-aided upper bound  from Theorem \ref{genie-mutual-info-bound.theorem}.

\begin{remark} \label{remark-detect}
It is interesting to notice that feedback errors make the residual interference behave as an
impulsive noise: it has very large variance with small probability $P_{e,{\rm fb}}$.
It is therefore clear that detecting the feedback errors and discarding the corresponding frames
yields significant improvements. Using this knowledge at the receiver (as in the rate bound
(\ref{genie-fberror-lb})), avoids the large  ``Jensen's penalty'' incurred by
the rate gap in (\ref{rate-fberror_general}), where the expectation with respect to the
feedback error events is taken {\em inside} the logarithm.
\hfill $\lozenge$
\end{remark}

\begin{remark}
We notice here that the naive ZF strategy examined in this paper is
{\em robust to feedback errors} in the following sense: the residual
interference experienced by a given UT depends only on that
particular UT feedback error probability. Therefore, a small number
of users with poor feedback channel quality (very high feedback
error probability) does not destroy the overall system performance.
This observation goes against the conventional wisdom that feedback
errors are ``catastrophic''.
\hfill $\lozenge$
\end{remark}

\subsection{Comparison between analog and digital channel feedback} \label{analog-digital-comp.sect}

Based upon the bounds developed in the previous subsections as well as the genie-aided upper bounds (computed
using Monte Carlo simulation) we can now compare analog, error-free digital, and QAM-based digital feedback.
Because the effect of downlink and common training is effectively the same for all feedback strategies,
we pursue this comparison under the assumption of perfect CSIR, i.e., perfect common and dedicated training corresponding
to $\beta_1=\beta_2 \rightarrow \infty$.
From \eq{AFGapAWGN} and \eq{rate-gap-digital-symbols} we have:
\begin{eqnarray} \label{analog-csir-compare}
\Delta R_{\textrm{CSIR}}^\textsc{AF} &\leq& \log \LB 1 +
\frac{1}{\beta_{\rm fb}} \RB \\
 \Delta R_{\textrm{CSIR}}^\textsc{DF} &\leq& \log \LB 1 + \frac{\frac{P}{N_0}}{\left (1 +
\frac{P}{N_0} \right)^{\beta_{\rm fb}}}  \RB
\end{eqnarray}
If $\beta_{\rm fb} = 1$ then analog and error-free digital feedback
both achieve essentially the same rate gap of $1$ bit per channel
user (per UT). However, if $\beta_{\rm fb} > 1$, the rate gap for
quantized feedback vanishes for $\frac{P}{N_0} \rightarrow \infty$.
This conclusion finds an appealing interpretation in the context of
rate-distortion theory. It is well-known (see for example
\cite{Gastpar} and references therein) that ``analog transmission''
(the source signal is input directly to the channel after suitable
power scaling) is an optimal strategy to send an i.i.d. Gaussian
source over a AWGN channel with the same bandwidth under quadratic
distortion. In our case, the source vector is $\hv_k$ (Gaussian and
i.i.d.) and the feedback channel is AWGN with with SNR
$\frac{P}{N_0}$. Hence, the fact that analog feedback cannot be
essentially outperformed for $\beta_{\rm fb} = 1$ is expected.
However, it is also well-known that if the channel bandwidth is
larger than the source bandwidth (which corresponds to the case
where a block of $M$ source coefficients are transmitted over
$\beta_{\rm fb} M$ channel uses with $\beta_{\rm fb} > 1$), then
analog transmission is strictly suboptimal with respect to a digital
scheme operating at the rate-distortion bound, because the
distortion with analog transmission  is $O((P/N_0)^{-1})$ whereas it
is  $O((P/N_0)^{-\beta_{\rm fb}})$ for digital transmission.

\begin{figure}[ht]
\begin{center}
\includegraphics[width=10cm]{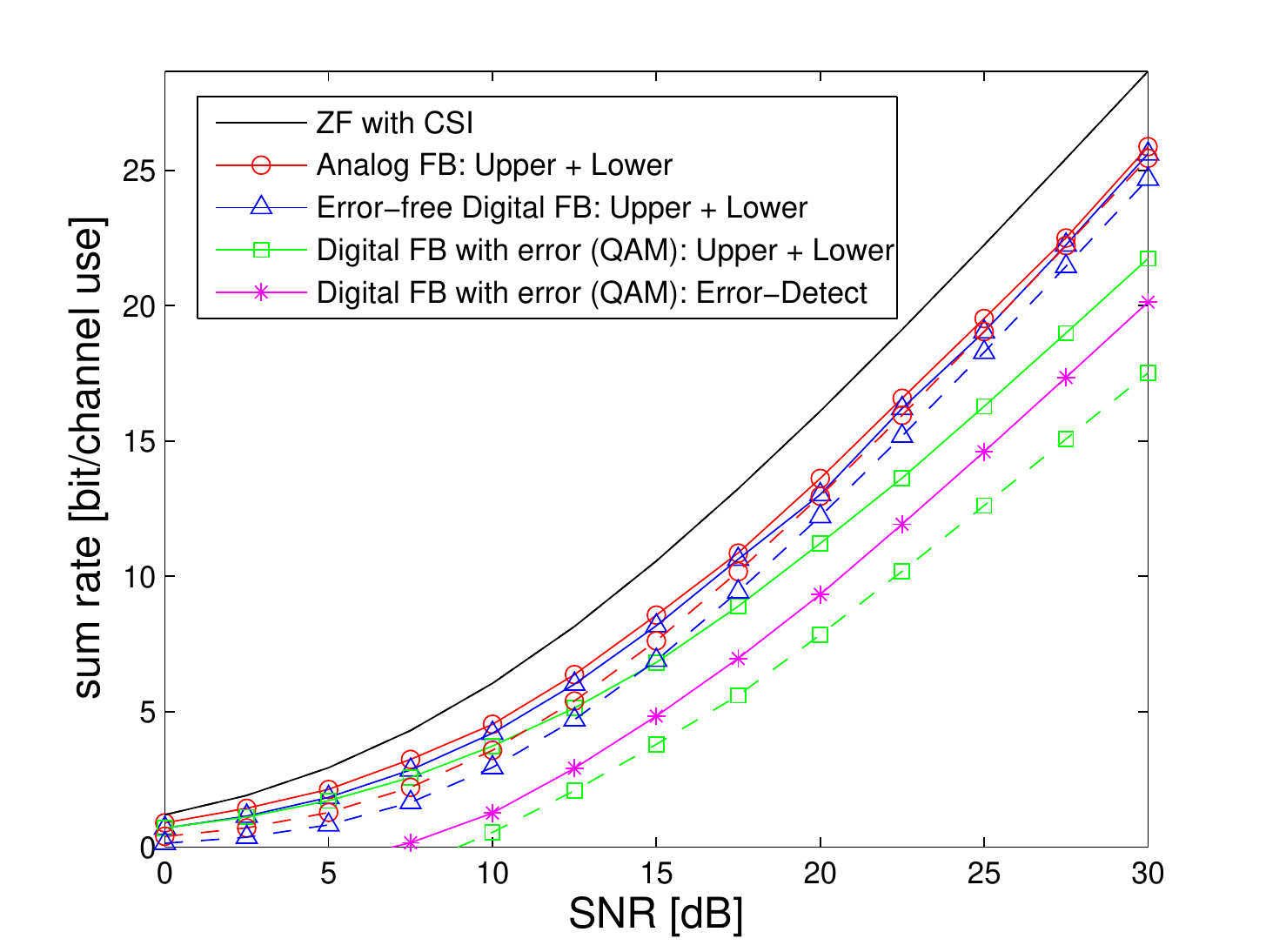}
\vspace{-0.5cm} \caption{Achievable rate lower (dotted lines) and
upper (solid lines) bounds for analog, error-free digital, and
QAM-based digital feedback for $M=4$ and $\beta_{\rm
fb}=1$.}\label{tput_snr_beta1}
\end{center}
\end{figure}

\begin{figure}[ht]
\begin{center}
\includegraphics[width=10cm]{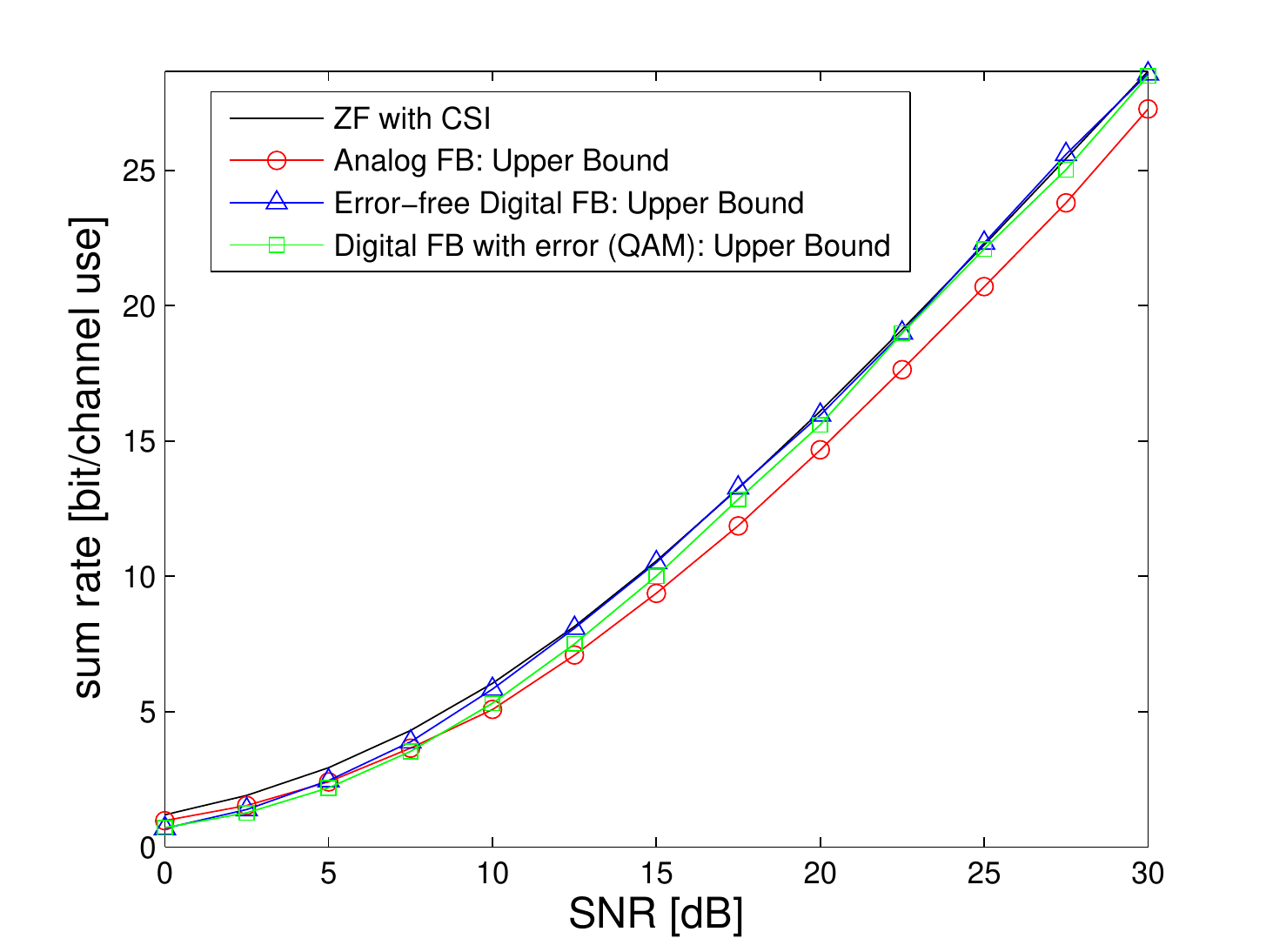}
\vspace{-0.5cm} \caption{Achievable rate upper bounds for analog,
error-free digital, and QAM-based digital feedback for $M=4$ and
$\beta_{\rm fb}=2$.}\label{tput_snr_beta2}
\end{center}
\end{figure}

\begin{figure}[ht]
\begin{center}
\includegraphics[width=10cm]{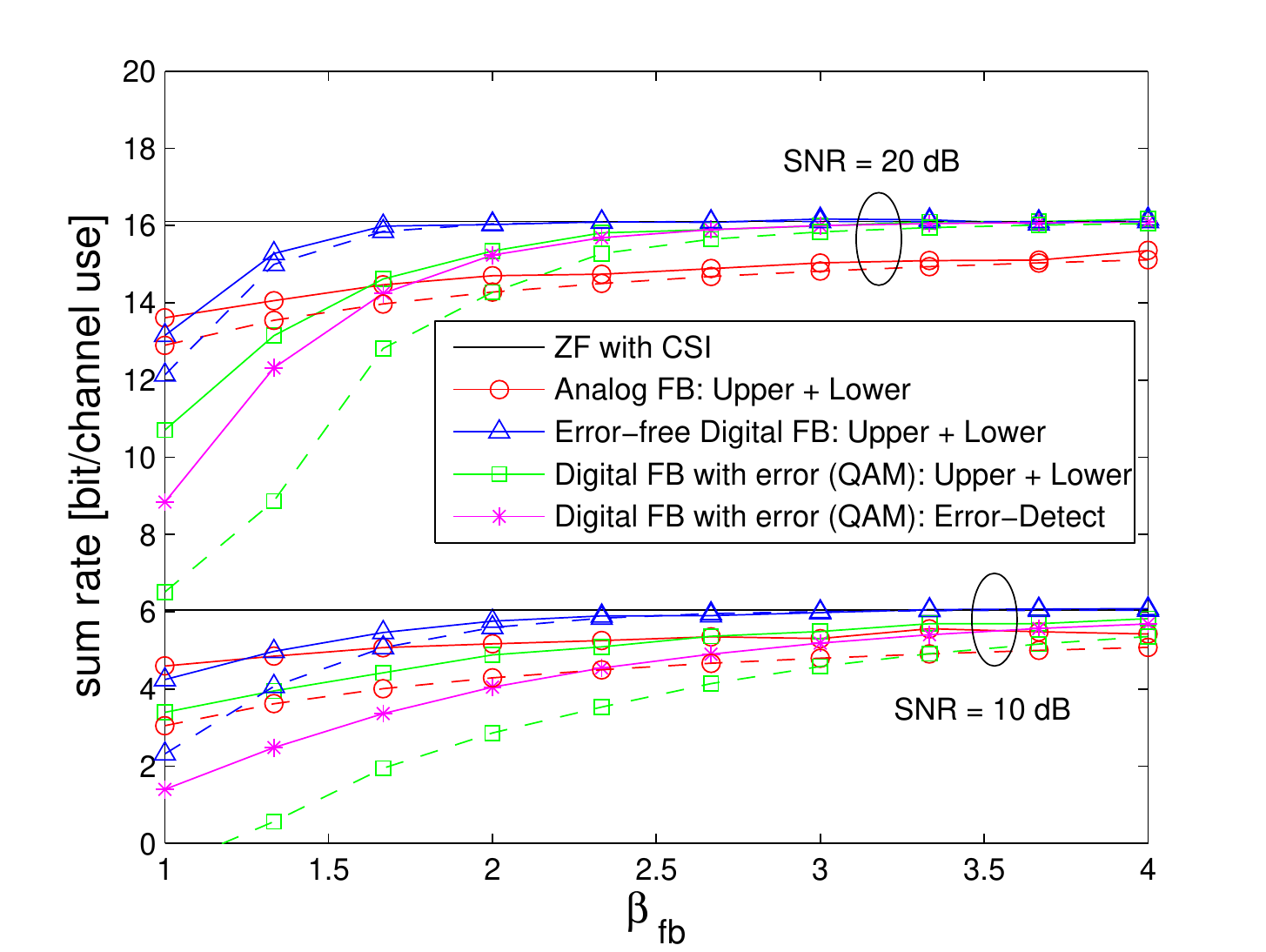}
\vspace{-0.5cm} \caption{Achievable rate lower (dotted lines) and
upper (solid lines) bounds for analog, error-free digital, and
QAM-based digital feedback for $M=4$ and $\frac{P}{N_0}= 10$ dB and
$\frac{P}{N_0}= 20$ dB.}\label{tput_beta}
\end{center}
\end{figure}

This conclusion is confirmed by the numerical results shown in Figures  \ref{tput_snr_beta1} and  \ref{tput_snr_beta2}. In Figure \ref{tput_snr_beta1} the
lower and genie-aided upper bounds are plotted for analog feedback, digital feedback without error, and digital feedback with error (QAM) versus SNR for
an $M=4$ system with $\beta_{\rm fb}=1$.  For digital feedback with error, the error detection bound in (\ref{genie-fberror-lb}) is also included. The
analog and error-free digital feedback schemes perform virtually identically and achieve a rate approximately $3$ dB away from the perfect channel state
information benchmark. Note also that the gap between the upper and lower bounds is not very large. For digital feedback with uncoded QAM~\footnote{These
results are obtained by optimizing the value of $1\leq \alpha \leq \beta_{\rm fb}$ for each SNR. We refer to this as ``envelope'', that is, the plotted
curve is the pointwise maximum of the rate vs. SNR curves for all $\alpha$.}, however, there is a substantial gap between the upper and lower bounds; this
gap and the performance with error detection is explained by Remark \ref{remark-detect}. In Figure \ref{tput_snr_beta2} only the genie-aided upper bounds
are plotted (because the lower and upper are nearly identical and thus are difficult to distinguish) for the same setting with $\beta_{\rm fb}=2$.  We see
that digital feedback with uncoded QAM outperforms analog feedback above approximately $5$ dB, and that the rate with digital feedback (with or without
errors) converges to the ideal rate as predicted earlier.  This figure confirms that the effect of feedback vanishes when digital feedback is used, with
or without errors, and $\beta_{\rm fb} > 1$. Finally, in Figure \ref{tput_beta} the bounds are plotted as a function of $\beta_{\rm fb}$ for fixed SNR
$\frac{P}{N_0} = 10$ dB and $\frac{P}{N_0} = 20$ dB. When $\beta_{\rm fb} \approx 1$ analog and error-free digital feedback are nearly equivalent, but as
$\beta_{\rm fb}$ is increased the rate with error-free digital quickly approaches the perfect channel state information rate.  When feedback errors are
introduced, digital feedback does eventually outperform analog and also approaches the ideal rate, but a larger $\beta_{\rm fb}$ is required.  It is also
worth noticing that as the SNR is increased, the value of $\beta_{\rm fb}$ at which digital (with or without errors) begins to outperform analog decreases
toward $1$: this is to expected based upon the fact that the effect of feedback vanishes as $\frac{P}{N_0} \rightarrow \infty$ for any $\beta_{\rm fb} >
1$ for digital, whereas it does not for analog feedback.

It is worth noting that the same basic conclusion, i.e., that digital feedback (with or without errors)
outperforms analog for sufficiently large $\beta_{\rm fb}$, also holds in the presence imperfect CSIR.
However, because imperfect CSIR leads to a residual term in the rate gap expression that does not vanish
(even for large $\frac{P}{N_0}$), the absolute difference between digital and analog feedback
is reduced.

\section{Channel state feedback over the MIMO-MAC} \label{sec:fading}

Orthogonal access in the feedback link requires $O(M^2)$ channel uses for the feedback, while the downlink capacity scales at best as $O(M)$. When the
number of antennas $M$ grows large, such a system would not scale well with $M$. On the other hand,  the inherent MIMO-MAC nature of the physical uplink
channel suggests an alternative approach, where multiple UT's simultaneously transmit on the MIMO uplink (feedback) channel and the spatial dimension is
exploited for channel state feedback too. This idea was considered for an FDD system in \cite{FastCSI} and analyzed in terms of the mean square error of
the channel estimate provided to the BS.

As in \cite{FastCSI}, we partition the $M$ users into $\frac{M}{L}$
groups of size $L$, and let UTs belonging to the same group
transmit their feedback signal simultaneously, in the same time frame. Each UT transmits its $M$ channel coefficients over $\beta_{\rm fb} M$
channel uses, with $\beta_{\rm fb} \geq 1$. Therefore, each group uses $\beta_{\rm fb} M$ channel symbols and the
total number of channel uses spent in the feedback is $\beta_{\rm fb} \frac{M^2}{L}$.
Choosing $L \propto M$ (e.g., $L = M/2$) yields
a total number of feedback channel uses that grows linearly with $M$,
such that the  feedback resource converges to a fixed fraction of the downlink capacity.
We assume that the uplink feedback channel is affected by i.i.d. block fading
(i.e., has the same distribution as the downlink channel) and that there is no feedback delay.

With respect to the analysis provided in \cite{FastCSI}, the present work differs in
a few important aspects: 1) we consider both analog and digital feedback; 2) although
our analog feedback model is essentially identical to the
FDD scheme of \cite{FastCSI}, we consider optimal MMSE estimation rather than Least-Squares estimation
(zero-forcing pseudo-inverse); 3) we put out results in the context of the rate gap framework,
that yields directly fundamental lower bounds on achievable rates,
rather than in terms of channel state estimation error.

\subsection{Analog Feedback} \label{sec-mimo-mac-analog}

In an analog feedback scheme, each UT feeds back a scaled noisy
version of its downlink channel, given by
$\frac{\sqrt{\beta_{\rm fb} P}}{\sqrt{\beta_1P + N_0}}\ {\sv_k}$ where
$\sv_k$ is the observation provided by the common training phase, defined in
(\ref{training-phase-1-rx}).
Due to the symmetry of the problem, we can focus on the simultaneous transmission of a single group of $L$ UTs.
Let $\Am =[\av_1 ~\cdots ~ \av_L] \in \CC^{M \times L}$ denote the uplink fading matrix for this
group of UTs (with i.i.d. entries, $\sim \Cc\Nc(0,1)$) and let for $k=1,\dots, L$
\begin{equation} \label{sucaminchia}
b_{k,j}  = \frac{\sqrt{\beta_{\rm fb} P}}{\sqrt{\beta_1P + N_0}} s_{k,j}
= \frac{\sqrt{\beta_{\rm fb}\beta_1} P}{\sqrt{\beta_1 P + N_0}} h_{k,j} +
\frac{\sqrt{\beta_{\rm fb} P}}{\sqrt{\beta_1P + N_0}} z_{k,j}
\end{equation}
denote the transmitted symbol by UT $k$ for its $j$-th channel coefficient,
where $s_{k,j}$ is the $j$-th component of $\sv_k$ and, from (\ref{training-phase-1-rx}), $z_{k,j}$ is
the common training AWGN.
For simplicity, we assume that the BS has perfect knowledge of the uplink channel state
$\Am$; we later consider the more general case and see that
the main conclusions are unchanged.

The $M$-dimensional received vector $\gv_j$, upon which the BS estimates the $j$-th antenna downlink channel coefficients
$h_{1,j}, \ldots, h_{L,j}$ of all users in the group, is given by:
\begin{eqnarray} \label{sucaminchia1}
\gv_j & = & \sum_{i=1}^L \av_i b_{i,j} + \widetilde{\wv}_j =
\Am \bv_j + \widetilde{\wv}_j
\end{eqnarray}
where $\widetilde{\wv}_j$ is an AWGN vector with i.i.d. elements $\sim \Cc\Nc(0,N_0)$.
From the i.i.d. jointly Gaussian statistics of the channel coefficients, downlink noise and uplink (feedback noise) it is
immediate to obtain the MMSE estimator for the downlink channel coefficient
$h_{k,j}$ in the form
\begin{equation} \label{ziopino1}
\widehat{h}_{k,j} =
c \ \av_k^\herm \left [  \beta_{\rm fb} P \Am\Am^\herm + N_0 \Id \right ]^{-1} \gv_j
\end{equation}
where we define the constant $c = \frac{\sqrt{\beta_{\rm fb}\beta_1}P}{\sqrt{\beta_1P + N_0}}$.
The corresponding MMSE, for given feedback channel matrix $\Am$, is given by
\begin{equation} \label{ziopino2}
\sigma_k^2(\Am) = 1 - c^2 \av_k^\her  \left [  \beta_{\rm fb} P \Am\Am^\herm + N_0 \Id \right ]^{-1} \av_k
\end{equation}

\begin{thm} \label{thm:analog_mimo_rate_gap}
If each UT feeds back its channel coefficients in analog fashion over the
MIMO MAC uplink channel, with groups of $L$ users simultaneously feeding back
and $\beta_{\rm fb} M$ channel uses per group,
the rate gap upper bound is given by:
\begin{equation} \label{shit-rategap}
\overline{\Delta R}^\textsc{AF}_\textsc{MIMO-MAC} = \log \LB 1 +
\frac{1}{M}\frac{\frac{P}{N_0}}{1 + \beta_2\frac{N_0}{P}} + \frac{M-1}{M}\ \frac{P}{N_0} \left [
\frac{1}{1 + \frac{\beta_1 P}{N_0}} + \frac{\frac{\beta_1 P}{N_0}}{1
+ \frac{\beta_1 P}{N_0}} {\sf mmse}\left (\frac{\beta_{\rm fb} P}{N_0}
\right ) \right ] \RB
\end{equation}
where we define the average channel state information estimation MMSE as
\begin{equation} \label{ziopino4}
{\sf mmse}(\rho) \triangleq \frac{1}{L} \sum_{k=1}^L \EE \left [ \frac{1}{1 + \rho \lambda_k} \right ]
\end{equation}
and where $\{\lambda_1,\ldots,\lambda_L\}$ denote the eigenvalues of the $L \times L$ central
Wishart matrix $\Am^\her \Am$.

Furthermore, if $L < M$ the rate gap is bounded and converges at high SNR to the constant
\begin{equation} \label{high-snr-rategap-fading}
\lim_{P/N_0 \rightarrow\infty} \overline{\Delta
R}^\textsc{AF}_\textsc{MIMO-MAC} = \log \LB 1 + \frac{1}{\beta_2
M} + \frac{M-1}{M} \LB \frac{1}{\beta_1} + \frac{1}{\beta_{\rm fb}(M-L)}  \RB \RB.
\end{equation}
\end{thm}
\vspace{5pt}
{\bf Proof:} See Appendix \ref{pf_mimo_mac_analog}. \QED

Comparing this expression to the rate gap for analog feedback over an AWGN channel
(\ref{analog-awgn-2}), we notice that an SNR (array) gain of $M-L$ is achieved (on the feedback channel)
when the feedback is performed over the MIMO MAC because the feedback (of $L$ users) is received over $M$
antennas.\footnote{At high SNR the feedback from a particular UT is effectively received
over an interference-free $1 \times (M - L + 1)$ channel because $L-1$ interfering signals are
nulled.  However, this results in only a $M-L$ multiplicative gain because $\EE[1/\chi^2_{2k}] = 1/(k-1)$.}
In addition, a factor of $L$ fewer feedback symbols are required when the feedback is performed
over the MIMO MAC ($\beta_{\rm fb} \frac{M^2}{L}$ vs. $\beta_{\rm fb}M^2$).
On the other hand, using the second line of the RHS of (\ref{ziopino5}) in Appendix \ref{proof_thm4}
it is immediate to show that for $L = M$ the rate gap upper bound grows unbounded as $\log \log \left (\frac{P}{N_0} \right )$.

From (\ref{high-snr-rategap-fading}) we can optimize the value of $L$ (assuming $L < M$)
for a fixed number of feedback channel uses, which we denote by $a M$ for some
$a \geq 2$ (if $L < M$ there must be at least two groups and thus we must have at
least $2M$ feedback symbols).
By letting $a M = \beta_{\rm fb} \frac{M^2}{L}$, we obtain $\beta_{\rm fb} = a \frac{L}{M}$. Using this in
(\ref{high-snr-rategap-fading}), we have that minimizing the rate gap bound
is equivalent to maximizing the term $L (M - L)$ for fixed $M$ and $L < M$. Therefore, the optimal group size
is given by  $L^* = \frac{M}{2}$. Substituting this value in (\ref{high-snr-rategap-fading}) yields
\begin{equation}\label{high-snr-rategap-fading2}
\log \LB 1 + \frac{1}{\beta_2
M} + \frac{2(M-1)}{M^2 \beta_{\rm fb}} + \frac{M-1}{M} \frac{1}{\beta_1} \RB
\end{equation}
and the corresponding total number of feedback symbols is $2 \beta_{\rm fb} M$.
Interestingly, we notice that in the regime of large $M$ the term that dominates the optimized rate gap bound
(\ref{high-snr-rategap-fading2}) corresponds to the downlink common training phase.
In fact, the terms corresponding to dedicated training and feedback vanish as $M$ increases.

When the total number of feedback symbols is larger or equal to $2M$ (i.e., $a \geq 2$)
numerical results verify that also at finite SNR the choice $L^* = \frac{M}{2}$ yields the best performance both in terms of
the achievable rate lower bound and of the the genie-aided upper bound. Hence,
the optimal MIMO-MAC feedback strategy is a combination of TDMA and SDMA.
In contrast, when total number of feedback symbols is strictly smaller
than $2M$ (i.e., $1 \leq a < 2$), choosing $L=M$ with $\beta_{\rm fb}=a$ is the
only option.   Although this choice yields an unbounded rate gap, it does provide reasonable performance
at finite SNR's.

A legitimate question at this point is the following: is the condition $L < M$ a fundamental limit of the
MIMO-MAC analog feedback in order to achieve a bounded rate gap, or is it due to the looseness of
Theorem \ref{rate-gap-bound.theorem}? In order to address this question, we examine the genie-aided
rate upper bound of Theorem \ref{genie-mutual-info-bound.theorem} and obtain the following rate upper bound:

\begin{thm}\label{thm:geniegapLM}
When a group of $L=M$ UTs feed back the channel coefficients simultaneously over $
\beta_{\rm fb}M$ channels uses of the fading MIMO-MAC, the difference between $R_k^{\rm ZF}$ and the
genie-aided upper bound
of Theorem \ref{genie-mutual-info-bound.theorem}  is uniformly bounded for all SNRs.
\end{thm}
\textbf{Proof } See Appendix \ref{proof_geniebound}. \hfill $\square$

Theorem \ref{thm:geniegapLM} suggests that if the UTs are able to obtain an estimate of their {\em instantaneous}
residual interference level in each frame, up to $M$ UTs can feedback their channel state information
at the same time. The ability of estimating the interference coefficients $\Ac_k$
(see (\ref{receiver-k}) and the comment following Theorem \ref{genie-mutual-info-bound.theorem}) depends critically
on the quality of the dedicated training. Hence, the dedicated training has a direct impact on the design and
efficiency of the channel state feedback. Such inter-dependencies between the different system components
can be illuminated thanks to the comprehensive system analysis carried out in this work
and are missed  by making  overly simplifying assumptions (e.g., genie-aided coherent detection with
perfect knowledge of the coefficients $\Ac_k$).

\begin{remark}
In \cite{FastCSI}, the same model in (\ref{sucaminchia}) for analog channel state feedback over the MIMO-MAC
uplink channel is considered. Instead of the linear MMSE estimator considered here, a zero-forcing approach (via the pseudo-inverse of the matrix $\Am$) is examined.   In the case of $L=M$ this yields an infinite
error variance, which does not make sense in light of the fact that each channel coefficient has unity variance.
This odd behavior can be avoided by performing an additional component-wise MMSE step.
As a matter of fact, performance very similar to what we have found for the full MMSE estimator can be obtained
for $L < M$ by using a zero-forcing receiver for the channel state feedback, followed by individual (componentwise) MMSE scaling.
We omit the analysis of such suboptimal scheme for the sake of brevity.
\hfill $\lozenge$
\end{remark}

\begin{remark}
It is also possible to analyze the more realistic scenario where the uplink channel matrix $\Am$ is known
imperfectly at the BS. We consider the following simple training-based scheme: the $L$ UTs within a feedback group
transmit a preamble of $\beta_{\rm up} L$ training symbols, where $\beta_{\rm up} \geq 1$ defines the
uplink training length (per UT).
Without repeating all steps in the details, the uplink channel $\Am$ admits the following decomposition:
\begin{equation} \label{uplink-ch-est}
\Am = \widehat{\Am} + \widetilde{\Am}
\end{equation}
where the channel estimate and estimation error $(\widehat{\Am}, \widetilde{\Am})$ are jointly
jointly Gaussian and independent, with per-component variances $1 - \sigma_{\rm up}^2$ and $\sigma_{\rm up}^2)$, respectively, with $\sigma_{\rm up}^2 = \frac{1}{1 + \beta_{\rm up} \frac{P}{N_0}}$.
Now, the MMSE estimation of the downlink channel coefficients $h_{k,j}$ is conditional with respect to
$\widehat{\Am}$. By repeating all previous steps, after a lengthy calculation that we do not report here for the sake of brevity,
we obtain the average estimation error in the form
\begin{eqnarray} \label{ziopino6}
\EE[\sigma_k^2(\widehat{\Am})] & = & \frac{1}{1 + \beta_1 \frac{P}{N_0}} +
\frac{\beta_1 \frac{P}{N_0}}{1 + \beta_1 \frac{P}{N_0}} \
{\sf mmse} \left (  \frac{\beta_{\rm up} \frac{P}{N_0}}{1 + \beta_{\rm up} \frac{P}{N_0} + L \beta_{\rm fb} \frac{P}{N_0}} \frac{\beta_{\rm fb}  P}{N_0}  \right )
\end{eqnarray}
where ${\sf mmse}(\cdot)$ was defined in (\ref{ziopino4}).
By comparing (\ref{ziopino6}) with (\ref{ziopino3}), we notice that they differ only in the argument of
the function ${\sf mmse}(\cdot)$.
The two expressions coincide for $\beta_{\rm up} \rightarrow \infty$, consistent with the fact that $\beta_{\rm up} \rightarrow \infty$ corresponds to perfect estimation of the channel matrix $\Am$.
Furthermore, for large SNR, the two arguments differ by  a constant multiplicative factor.
Hence, apart from this constant factor that depends on the uplink training parameter $\beta_{\rm up}$,
the conclusions about the rate gap obtained for the case of perfect uplink channel knowledge also hold
for the case of  training-based uplink channel estimation.
\hfill $\lozenge$
\end{remark}

\subsection{Digital Feedback}

In the case of digital feedback, we let $L \leq M$ UTs multiplex their channel state
feedback codewords at the same time.
The resulting MIMO-MAC channel model is again given by (\ref{sucaminchia1}),
but now the vector $\bv_j$ contains the $j$-th symbols of the
feedback codewords of the $L$ UTs sharing the same feedback frame.
As in Section \ref{digital.sect}, we assume that feedback messages of $\alpha (M-1) \log_2 \frac{P}{N_0}$ bits are sent in
$\beta_{\rm fb} (M-1)$ channel uses. Hence, the feedback symbols transmitted by the $L$ UT's can be grouped in
a $L \times \beta_{\rm fb} (M-1)$ matrix, while the BS has an
$M \times \beta_{\rm fb} (M-1)$ observation upon which to estimate the
transmitted symbols.   We again assume each feedback symbol has average energy $P$.

Suppose that the BS receiver operates optimally, by using a joint ML decoder for all the simultaneously transmitting
UTs. The high-SNR error probability performance of the MIMO-MAC channel was characterized in terms of the
diversity-multiplexing tradeoff in \cite{tse2004dmt}. In particular, when each user transmits at rate $r \log_2 \frac{P}{N_0}$ bits/symbol (i.e., with {\em multiplexing gain} $r$) over the MIMO-MAC with i.i.d. channel fading (as considered here), the optimal ML decoder achieves an individual user average error probability
\[ P_{e,{\rm fb}} \; \doteq \; \left ( \frac{P}{N_0} \right )^{-d^*(r)} \]
where the ``dot-equality'' notation, introduced in \cite{zheng2003dam,tse2004dmt},
indicates that $\lim_{P/N_0 \rightarrow \infty} \frac{-\log P_{e,{\rm fb}}}{\log P/N_0} = d^*(r)$.
The error probability SNR exponent $d^*(r)$ is referred to as the optimal {\em diversity gain} of the system.
Particularizing the results of \cite{tse2004dmt} to the case of $L \leq M$ users with 1 antenna each,
transmitting to a receiver with $M$ antennas, the optimal diversity gain is given by
\begin{equation} \label{babaciu1}
d^*(r) = \left \{ \begin{array}{ll}
M (1 - r) & \mbox{for} \;\; 0 \leq r \leq 1 \\
0 & \mbox{otherwise} \end{array} \right .
\end{equation}
This is the same exponent of a channel with a single user with a
single antenna, transmitting to a receiver with $M$ antennas
(single-input multiple-output with receiver antenna diversity). In
other words, under our system parameters, each UT achieves an error
probability that decays with SNR as if TDMA on the feedback link was
used (as if the UT transmitted its feedback message alone on the
MIMO uplink channel). From what is said above, it follows that the
multiplexing gain of all UTs is given by $r =
\frac{\alpha}{\beta_{\rm fb}}$. Furthermore, from the derivation of
Section \ref{errors.sect}, we require that $1 < \alpha < \beta_{\rm
fb}$. It follows that the average feedback error message probability
in the MIMO-MAC fading channel is given by
\begin{equation} \label{babaciu2}
P_{e, {\rm fb}} = \left ( \frac{P}{N_0} \right )^{-M (1 - \alpha/\beta_{\rm fb})} \times g \left (\frac{P}{N_0} \right )
\end{equation}
where $g(x)$ is some sub-polynomial function, such that $\lim_{x \rightarrow \infty} x^{-\epsilon} g(x) = 0$ for all
fixed $\epsilon > 0$.

If we examine the rate-gap expression with digital feedback
(\ref{rate-fberror_general}), we see that in order to achieve a bounded rate
gap the error probability $P_{e,{\rm fb}}$ must go to zero at least as fast as
$\left ( \frac{P}{N_0} \right )^{-1}$. From (\ref{babaciu2}) we have that for all $1 < \alpha < \beta_{\rm fb}$
such that $M(1 - \alpha/\beta_{\rm fb})$ is strictly larger than 1,
the resulting rate gap is bounded and the effect of feedback errors vanishes.
This imposes the condition $\beta_{\rm fb} >  \frac{M}{M-1}$ and
$\alpha < \frac{M-1}{M} \beta_{\rm fb}$, which is stricter than the condition
 $\beta_{\rm fb} >  1$ and $\alpha < \beta_{\rm fb}$ needed in the case of TDMA an unfaded feedback channel
 previously analyzed in Section \ref{errors.sect}.

We conclude that a bounded rate gap can also be achieved with digital feedback on the MIMO-MAC uplink channel.
Therefore, also in this case we can achieve a number of feedback channel uses that scales linearly with the number of the BS antennas $M$. Explicit design of codes that achieve the optimal divesity-multiplexing tradeoff of MIMO-MAC channels
is not an easy task in general. In the particular case of $M$ users with one antenna each,
simple explicit constructions of MIMO-MAC codes for the digital channel state feedback are presented
\cite{raj-ISIT09}.
These codes can be optimally decoded by using a Sphere Decoder \cite{viterbo1999ulc, damen2003mld}
and achieve the performance promised by the above analysis.
It should be noticed, however, that while in the AWGN case the term $\frac{P}{N_0} P_{e,{\rm fb}}$
in the rate gap expression vanishes rapidly (faster than polynomially, in $P/N_0$),
in the MIMO-MAC fading case it vanishes only as $(P/N_0)^{1 - M(1 - \alpha/\beta_{\rm fb})}$.
Thus, for finite SNR the rate gap may be significantly larger than in the case of unfaded
feedback channel and the optimal tradeoff between quantization distortion and the feedback error probability must be sought
by careful optimization of the parameters $\alpha$ and $\beta_{\rm fb}$ (see details in  \cite{kumar2009itw}).
Also, the same observations about detecting feedback errors at the UTs and discarding the corresponding frames
made at the end of Section \ref{errors.sect} apply here.


\subsection{Numerical example}

\begin{figure}[t]
\begin{center}
\includegraphics[width=10cm]{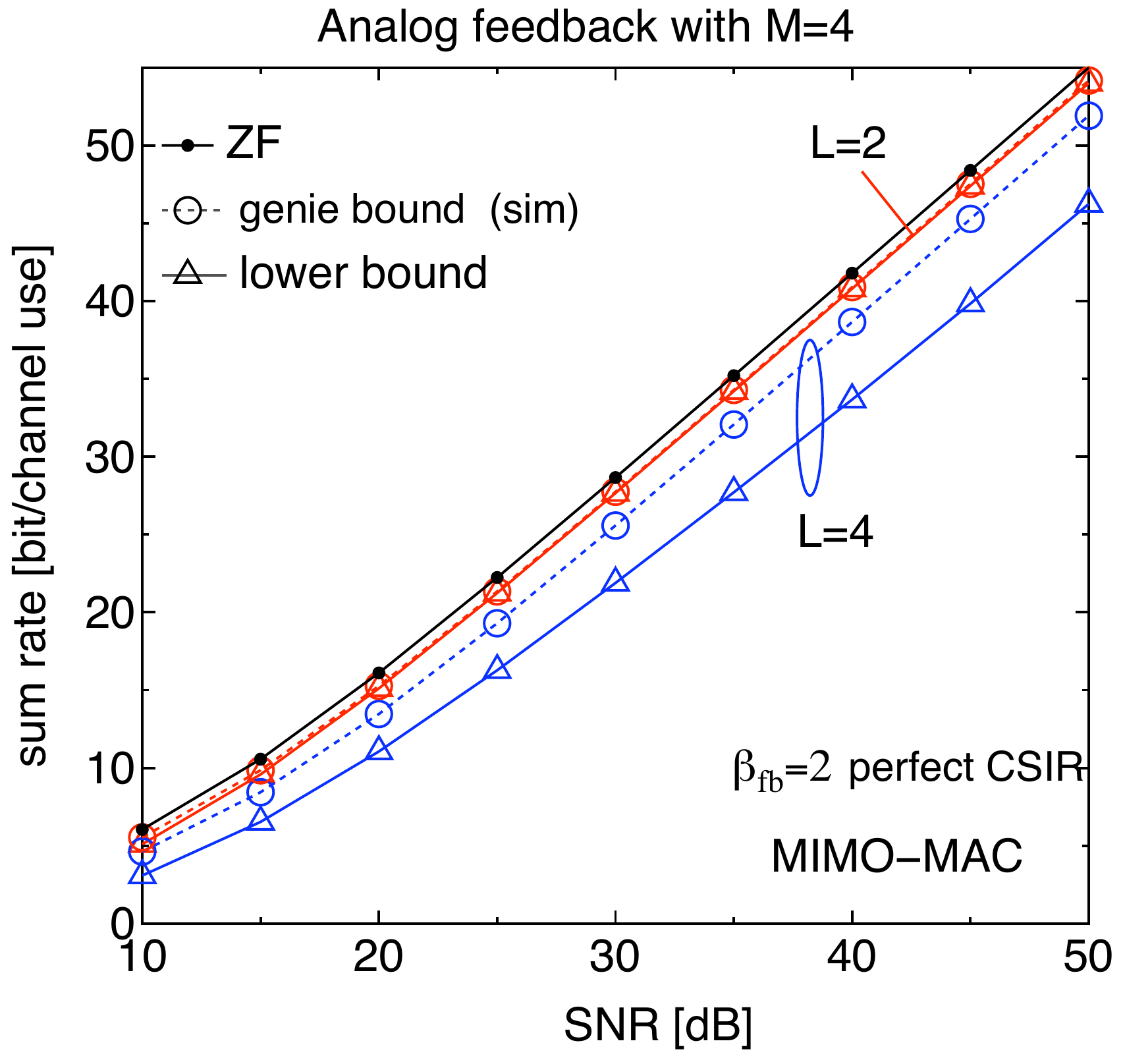}
\end{center}
\vspace{-0.5cm} \caption{Impact of $L$ with analog feedback over
MIMO-MAC} \label{ImpactLwAnaM4}
\end{figure}

Fig.~\ref{ImpactLwAnaM4} shows both the genie-aided upper bound of
Theorem \ref{genie-mutual-info-bound.theorem} and the lower bound
based on (\ref{shit-rategap}) of analog feedback over a fading
MIMO-MAC for $M=4$ and $L=2,4$. We assume perfect CSIR.
We notice that for $L=2$, the lower bound coincides with the genie-aided upper
bound and comes very close to the performance of ZF with ideal
CSIT. For $L = M$, the rate gap of the lower bound
(\ref{shit-rategap}) is  unbounded but the double logarithmic growth
$\log\log(P/N_0)$ yields a very small gap for a wide range of practical SNRs.
The genie-aided bound achieves a constant rate gap even for $L=M$, in accordance with
Theorem \ref{thm:geniegapLM}.
Although not shown here, a system using $M=4,L=2$ and $\beta_{\rm fb}=1$
does outperform $M=L=4$, $\beta_{\rm fb}=2$ (both configurations use a total of $8$
feedback symbols per frame) in terms of the lower bound and the genie-aided upper bound throughout
the SNR range shown; this validates our earlier claim about the optimality of $L = \frac{M}{2}$ whenever at least
$2M$ feedback symbols are used.

\begin{figure}[t]
\begin{center}
\includegraphics[width=10cm]{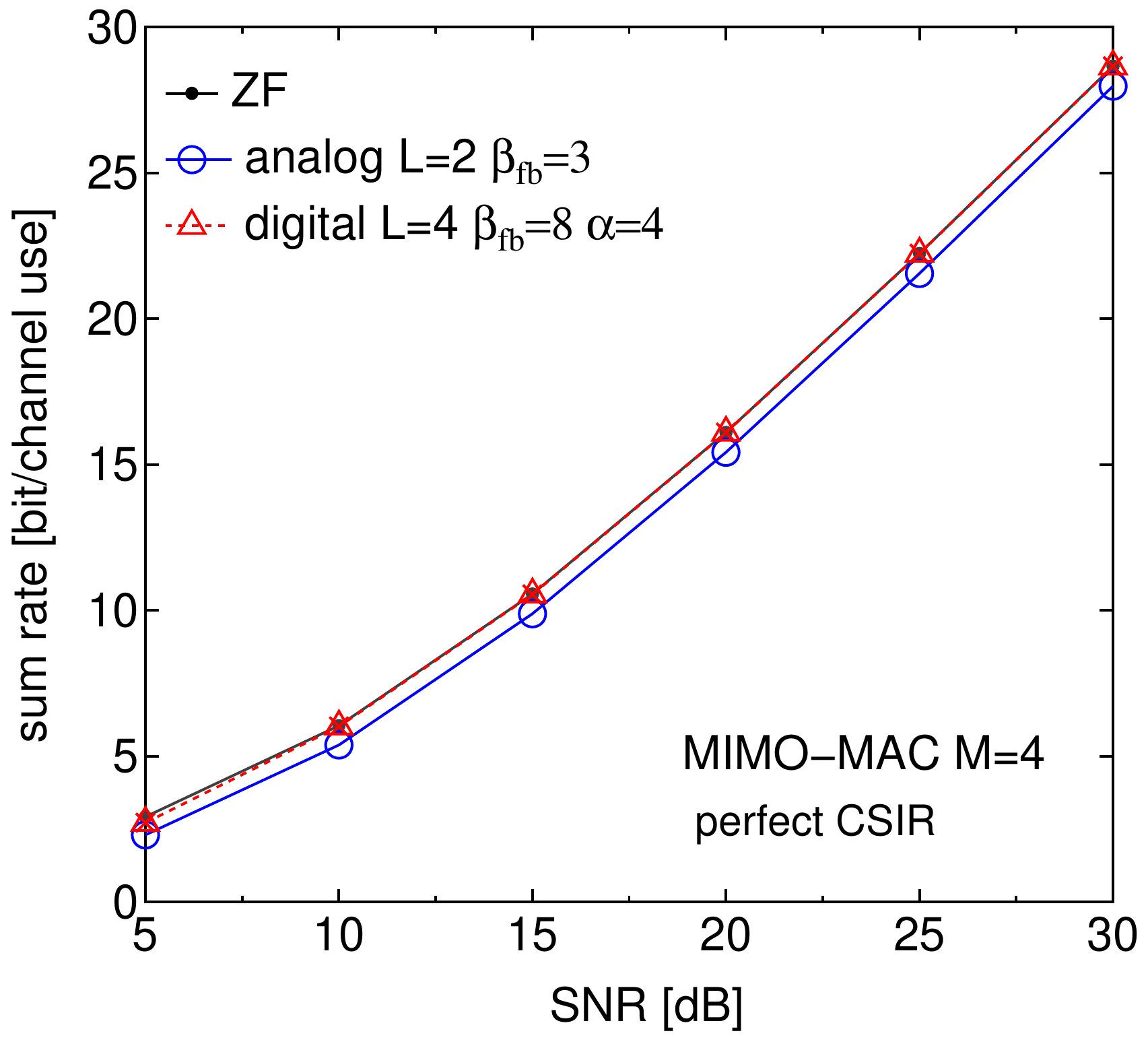}
\vspace{-0.5cm} \caption{Achievable rate lower bounds for analog and
digital feedback for $M=4$ and $24$ total feedback symbols.}
\label{mimo-mac_analog_dig}
\end{center}
\end{figure}

Fig. \ref{mimo-mac_analog_dig} compares the achievable rates of analog and digital feedback
schemes based on the rate gap (50), (42), over a fading MIMO-MAC for
$M=4$. For the digital feedback we assume that there exist some code
achieving the outage probability (57) with $g(P/N_0)=1$. We compare both schemes for
the same total amount of the feedback symbols (24 symbols). For the
analog feedback we choose $L=2, \beta_{\rm fb}=3$,  while for the
digital feedback we let $L=4, \beta_{\rm fb}=8, \alpha=4$. We
observe that the digital feedback achieves near-optimal sum rate
over the all SNR ranges while the analog feedback achieves a
constant gap of roughly 0.7 bit/channel use. Surprisingly, the
digital feedback is able to let M users transmit simultaneously
while vanishing both the quantization error and the feedback error.

\section{Effects of CSIT feedback delay} \label{feedback-delay.sect}

In this section we wish to take into account the effect of feedback
delay in a setting where the fading is temporally correlated. We
assume that the fading is constant within each frame, but changes
from frame to frame according to a stationary random process. In
particular, assuming spatial independence, each entry of $\hv_k$
evolves independently according to the same complex circularly
symmetric Gaussian stationary ergodic random process, denoted by
$\{h(t)\}$, with mean zero, unit variance and power spectral density
(Doppler spectrum) denoted by $S_h(\xi)$, $\xi \in [-1/2,1/2]$, and
satisfying $\int_{-1/2}^{1/2} S_h(\xi) d\xi = 1$, Notice that the
discrete-time process $\{h(t)\}$ has time that ticks at the frame
rate.

Because of symmetry and spatial independence,  we can neglect the UT index $k$ and the antenna index
and consider scalar rather than vector processes. Generalizing (\ref{training-phase-1-rx}), the observation available at each UT at  time $t - d$
from the common training phase takes on the form
\begin{equation} \label{training-phase-1-rx-correlation}
\left \{ s(t - \tau) = \sqrt{\beta_1 P} h(t - \tau) + z(t - \tau) : \tau = d,d+1,d+2,\ldots, \infty \right \}
\end{equation}
where $d$ indicates the feedback delay in frames. This means that
the channel state feedback to be used by the BS at frame time $t$ is
formed from noisy observations of the channel up to time $t-d$.
We consider a scheme where each UT at frame $t-d$ produces the MMSE estimate of its channel at
frame $t$ and sends this estimate (using either analog or digital
feedback) to the BS; the BS uses the received feedback to choose the beamforming
vectors used for data transmission in frame $t$.

\subsection{Estimation Error at UT}

The key quantity in the associated rate gap is the MMSE prediction error at the UT.
Let $\widetilde{h}(t)$ denote the MMSE estimate of $h(t)$ given the observations in
(\ref{training-phase-1-rx-correlation}). Given the joint Gaussianity of $h$ and $s$, we can write
\begin{equation} \label{ziobeduino}
h(t) = \widetilde{h}(t) + n(t)
\end{equation}
where $\EE[|n(t)|^2] = \sigma_1^2$ is the estimation MMSE,
and $\widetilde{h}(t)$ and $n(t)$ are independent with
$\EE[|\widetilde{h}(t)|^2] = 1 - \sigma_1^2$. From classical Wiener
filtering theory \cite{poor-book}, the one-step prediction ($d = 1$)
MMSE error is given by
\begin{equation} \label{noisy-prediction}
\epsilon_1(\delta) = \exp\left ( \int_{-1/2}^{1/2} \log(\delta + S_h(\xi)) d\xi \right ) - \delta
\end{equation}
where $\delta= N_0 / (\beta_1 P)$ is the observation noise variance.
The filtering MMSE ($d = 0$) is related to $\epsilon_1(\delta)$ as
\begin{equation} \label{noisy-filtering}
\epsilon_0(\delta) = \frac{\delta \epsilon_1(\delta)}{\delta + \epsilon_1(\delta)}.
\end{equation}
The scenario considered in all previous sections corresponds to i.i.d. fading (across blocks) and $d=0$,
in which case $\epsilon_1(\delta) = 1$ (past observations are useless) and thus
$\sigma_1^2 = \epsilon_0(\delta) = (1 + \beta_1 \frac{P}{N_0})^{-1}$, which coincides
with (\ref{estimation-error-common-phase}). More in general, in this section we shall
consider~\footnote{We focus on the case $d = 1$, because it is very relevant in practical
applications. For example, high-data rate downlink systems such as 1xEv-Do \cite{HDR} already implement
a very fast channel state feedback with at most one frame delay. Furthermore,
the one-step prediction case allows an elegant closed-form analysis.}
$\sigma_1^2 = \epsilon_d(\delta)$ for $d = 0, 1$.

We distinguish two cases of channel fading statistics:
{\em Doppler process} and {\em regular process}:
\begin{itemize}
\item  We say that $\{h(t)\}$ is a Doppler process if $S_h(\xi)$ is strictly
band-limited to $[-F,F]$, where $F < 1/2$ is the maximum Doppler
frequency shift, given by $F = \frac{vf_c}{c} T_f$, where $v$ is the
mobile terminal speed (m/s), $f_c$ is the carrier frequency (Hz),
$c$ is light speed (m/s) and $T_f$ is the frame duration (s) \cite{biglieri1998fci}.
A Doppler process satisfies $\int_{-F}^F \log
S_h(\xi) d\xi > -\infty$,
and has prediction error\footnote{As in \cite{lapidoth2005acs}, the same result holds for
a wider class of processes such that the Lebesgue measure of the set
$\{\xi \in [-1/2,1/2] : S_h(\xi) = 0\}$ is equal to $1 - 2F$, and such that $\int_{\Dc} \log (S_h(\xi)) d\xi > -\infty$
where $\Dc$ is the support of $S_h(\xi)$.}
\begin{equation} \label{e1-doppler}
\epsilon_1(\delta) = \delta^{1 - 2F} \exp\left ( \int_{-F}^{F} \log(\delta + S_h(\xi)) d\xi \right ) - \delta
\end{equation}
Therefore, $\lim_{\delta \rightarrow 0} \epsilon_1(\delta) = 0$.
\item
We say that $\{h(t)\}$ is a regular process if $\epsilon_1(0) > 0$
(see \cite{lapidoth2005acs} and references therein). In
particular, a process satisfying the Paley-Wiener condition
\cite{poor-book} $\int_{-1/2}^{1/2} \log S_h(\xi) d\xi > -\infty$ is regular.
\end{itemize}

For the case of no delay ($d=0$), for either type of process the estimation error goes to zero
with the observation noise, i.e., $\epsilon_0(\delta) \rightarrow 0$ as $\delta \rightarrow 0$.
However, they differ sharply in terms of prediction error: $\epsilon_1(\delta)$ is strictly positive for
a regular process (even as $\delta \rightarrow 0$), whereas $\epsilon_1(\delta) \rightarrow 0$
for Doppler processes as quantified in the following:
\begin{lemma} \label{prediction.lemma}
The noisy prediction error of a Doppler process satisfies
\begin{equation}
\epsilon_1(\delta) = \kappa \delta^{1 - 2F} + O(\delta)
\end{equation}
for $\delta \downarrow 0$, where $\kappa$ is a constant term independent of $\delta$.
\end{lemma}

{\bf Proof:}
Applying Jensen's inequality to \eq{e1-doppler} from the fact that $\int S_h(\xi) d\xi = 1$, we obtain the upper bound
\begin{equation} \label{termP-1}
\epsilon_1\left ( \delta \right ) \leq \delta^{1-2F} \left [ \left (\frac{1}{2F} + \delta  \right )^{2F} - \delta^{2F} \right ]
\end{equation}
Using the fact that $\log$ is increasing, we arrive at the lower bound
\begin{equation} \label{termP-2}
\epsilon_1\left (\delta \right ) \geq \delta^{1-2F} \left [ \exp\left (\int_{-F}^F \log S_h(\xi) d\xi\right ) - \delta^{2F} \right ]
\end{equation}
Combining these bounds we obtain the result. \QED

\subsection{Rate Gap Upper Bound}

When analog feedback is used, each UT transmits a scaled version of its MMSE
estimate $\widetilde{h}(t)$ over the feedback channel. The only difference from the scenarios studied in
Sections \ref{sec-mimo-mac-analog} (AWGN feedback channel) and \ref{analog.sect} (MIMO MAC feedback channel)
is that the estimation error at the UT is $\epsilon_d\LB N_0 / (\beta_1 P) \RB$ rather than
$(1 + \beta_1 P/N_0)^{-1}$.
As a result, a simple calculation confirms that the expressions for the rate gap upper bound given
in Theorems \ref{thm:analog_rate_gap} (AWGN) and \ref{thm:analog_mimo_rate_gap} (MIMO-MAC)
apply to the present if $\epsilon_d\LB N_0 / (\beta_1 P) \RB$ is substituted for $(1 + \beta_1 P/N_0)^{-1}$.
The same equivalence holds for digital feedback: each UT quantizes its MMSE estimate $\widetilde{h}(t)$,
and as a result the rate gap upper bound given in Theorem \ref{thm:digital_rate_gap} applies with the
same substitution. For the sake of brevity, the expressions for the rate gap upper bound are not provided here.

In fact, the effect of feedback delay is most clearly illustrated by considering perfect feedback
(i.e., $\beta_{\rm fb} \rightarrow \infty$), in which case (at frame $t$) the BS has perfect knowledge of
$\widetilde{h}(t)$, the UT's prediction of $h(t)$ based on common training observations up to frame $t-d$.
For the sake of simplicity we further assume perfect dedicated training (i.e., $\beta_2 \rightarrow \infty$),
in which case the rate gap upper bound is
\begin{equation} \label{rategap_predict}
\overline{\Delta R}^\textsc{predict} =
\log \left ( 1 + \frac{P}{N_0} \frac{M-1}{M} \epsilon_d \LB \frac{N_0}{\beta_1 P} \RB  \right).
\end{equation}

We now analyze the cases of no delay and one-step delay for both types of processes.
\paragraph{No feedback delay ($d=0$)}
Because using past observations can only help, the filtering error is no larger than
the error if the past is ignored, i.e., $\epsilon_0\left ( \delta \right ) \leq (1 + \beta_1 P/N_0)^{-1}$.
It thus follows that for both Doppler and regular processes the rate gap is bounded.
Based upon (\ref{noisy-filtering}), Lemma \ref{prediction.lemma}, and the property
$\epsilon_0 (0) > 0$ for regular processes, it is straightforward to see that $(P/N_0) \epsilon_0\left ( N_0/(\beta_1 P)  \right ) \rightarrow
\frac{1}{\beta_1}$ as $\frac{P}{N_0} \rightarrow \infty$ for either regular or Doppler processes.
As a result, the rate gap upper bound in (\ref{rategap_predict}) converges to $\log (1 + 1 / \beta_1)$ at high SNR.
This matches the high SNR expression for block-by-block estimation in (\ref{AFGapAWGN}), showing that
filtering does not provide a significant advantage at asymptotically high SNR.
However, as later illustrated through numerical results, this convergence occurs extremely slowly for Doppler
processes or highly correlated regular processes, in which case filtering does provide a non-negligible gain
over a wide range of SNR's.


\paragraph{Feedback delay ($d=1$)}
For regular fading process, since $\epsilon_1(0) > 0$, the quantity
$(P/N_0) \epsilon_1 \left ( N_0/(\beta_1 P) \right )$ increases
linearly with $\frac{P}{N_0}$ and thus the rate gap upper bound
$\overline{\Delta R}^\textsc{predict}$ grows like $\log
\frac{P}{N_0}$. As a result, the achievable rate lower bound
$R_k^{\textrm ZF} - \overline{\Delta R}^\textsc{predict}$ is bounded
even as $P/N_0 \rightarrow \infty$. In addition, in Appendix
\ref{app-bounded} we show that the genie-aided upper bound is also
bounded due to the fundamentally non-deterministic nature of regular
processes. This shows that with delayed feedback and a channel that
evolves according to a regular fading process, a system that makes
use of zero-forcing naive beamforming  to $M$ users becomes
interference limited. \footnote{In order to have a non-interference
limited system we can always use TDMA and serve one user at a time.
However, in this case the sum-rate would grow like $\log(P/N_0)$
instead of $M \log(P/N_0)$ as promised by the MIMO downlink with
perfect CSIT.} This behavior holds even with CSIR (i.e., letting
$\beta_1 \rightarrow \infty$).

Fortunately, physically meaningful fading processes belong to the class of Doppler processes, at least over a time-span where they can be considered stationary. For a practical relative speed between BS and UT, such time span is much
larger than any reasonable coding block length. Hence, we may say that Doppler processes are more the rule than the exception. In this case, the system behavior is radically different.
Using Lemma \ref{prediction.lemma}, at high SNR the rate gap upper bound is
\begin{equation}
\log \left ( 1 + \frac{M-1}{M} \frac{P}{N_0}  \left( \kappa \left( \frac{\beta_1 P}{N_0}\right)^{2F-1}
+ O\left (  \frac{N_0}{\beta_1 P}\right ) \right) \right),
\end{equation}
and thus the rate gap grows like $2 F \log \frac{P}{N_0}$. Using
this in the rate lower bound of Corollary (\ref{cor-lower}), and
considering the pre-log factor in high-SNR, we have that the system
sum-rate is lowerbounded by
\begin{equation} \label{pre-log}
\sum_{k=1}^M R_k \geq M (1 - 2F) \log \frac{P}{N_0} + O(1)
\end{equation}
This shows that a multiplexing gain of $M(1 - 2F)$ is achievable.

\begin{remark}
If perfect CSIR is assumed, an interesting singularity is observed for Doppler processes.
Under this assumption each UT is able to perform \emph{perfect prediction} of its channel state
on the basis of its past noiseless observations of the channel, by the definition of a Doppler
process.  Thus, it is as if there is no delay and the full multiplexing gain of $M$ is
achieved (even if the feedback link is imperfect).
On the other hand, if perfect CSIR is not assumed and UT's learn their
channel through $\beta_1 M$ common training symbols, for any finite value of $\beta_1$ a multiplexing
gain of only $M(1-2F)$ is achieved.  This point illustrates, again, that neglecting some system aspects
may yield to erroneous conclusions. In this case, by properly modeling imperfect CSIR we have illuminated the
impact of the UTs speed (which determines the channel Doppler bandwidth ) on the system
achievable rates in a concise and elegant way.
\hfill $\lozenge$
\end{remark}

\begin{remark}
It is interesting to notice here the parallel with the results of
\cite{lapidoth2005acs} on the high-SNR capacity of the single-user
scalar ergodic stationary fading channel with no CSIR and no CSIT,
where it is shown that for a class of {\em non-regular} processes
that includes the Doppler processes defined here, the high-SNR
capacity grows like $\Lc \log (P/N_0)$, where $\Lc$ is the Lebesgue
measure of the set $\{\xi \in [-1/2,1/2] : S_h(\xi) = 0\}$.
In our case, it is clear that $\Lc = 1 - 2F$. These results, as ours, rely on the
behavior of the noisy prediction error $\epsilon_1(\delta)$ for small $\delta$.
\hfill $\lozenge$
\end{remark}

\subsection{Examples}

We now present numerical results for the Jake's model and the Gauss-Markov model,
which are two widely used Doppler and regular processes, respectively.
The classical Jakes' correlation model has the following spectrum \cite{tse2005fwc,goldsmith2005wc}
\begin{equation}
S_h(\xi) = \frac{1}{\pi\sqrt{F^2-\xi^2}}, \;\;\; -F \leq \xi \leq F,
\end{equation}
and auto-correlation function $J_0(2 \pi F \tau)$.
No closed-form solution is known for the prediction or filtering error.
Under the Gauss-Markov model (i.e., auto regressive of order 1) the channel evolves in time as:
\begin{equation} \label{eqn:evol}
h(\tau) = r h(\tau-1) + \sqrt{1-r^2} \Delta(\tau)
\end{equation}
where $r$ is the correlation coefficient ($0<r<1$) and
the innovation process $\Delta(\tau)$ is unit-variance complex Gaussian, i.i.d. in time.
The prediction error for such model can be written in closed-form and is given by
(see for example \cite{mari-jsac})
\begin{eqnarray}
    \epsilon_1(\delta) &=& (1-r^2)\left[1
    +\frac{-(1+\delta)+\sqrt{1+\delta^2+2\delta
    \frac{1+r^2}{1-r^2}}}{2}\right]
\end{eqnarray}
For the Jakes' model we have $F = \frac{v f_c}{c}T_f$.  In all results we consider
$f_c=2$ GHz and $T_f = 1$ msec.  Motivated by the maximum-entropy principle \cite{cover-book},
several works in wireless communication modeled channel fading as  Gauss-Markov
process with one-step correlation coefficient  $r=J_0(2 \pi F)$, given by Jakes' model.
Comparing the performance of the true Jakes' model with its Gauss-Markov
maximum-entropy approximation, we will point out that the latter may be overly pessimistic for high-speed
mobile terminals.

\begin{figure}[t]
\begin{center}
\includegraphics[width=10cm]{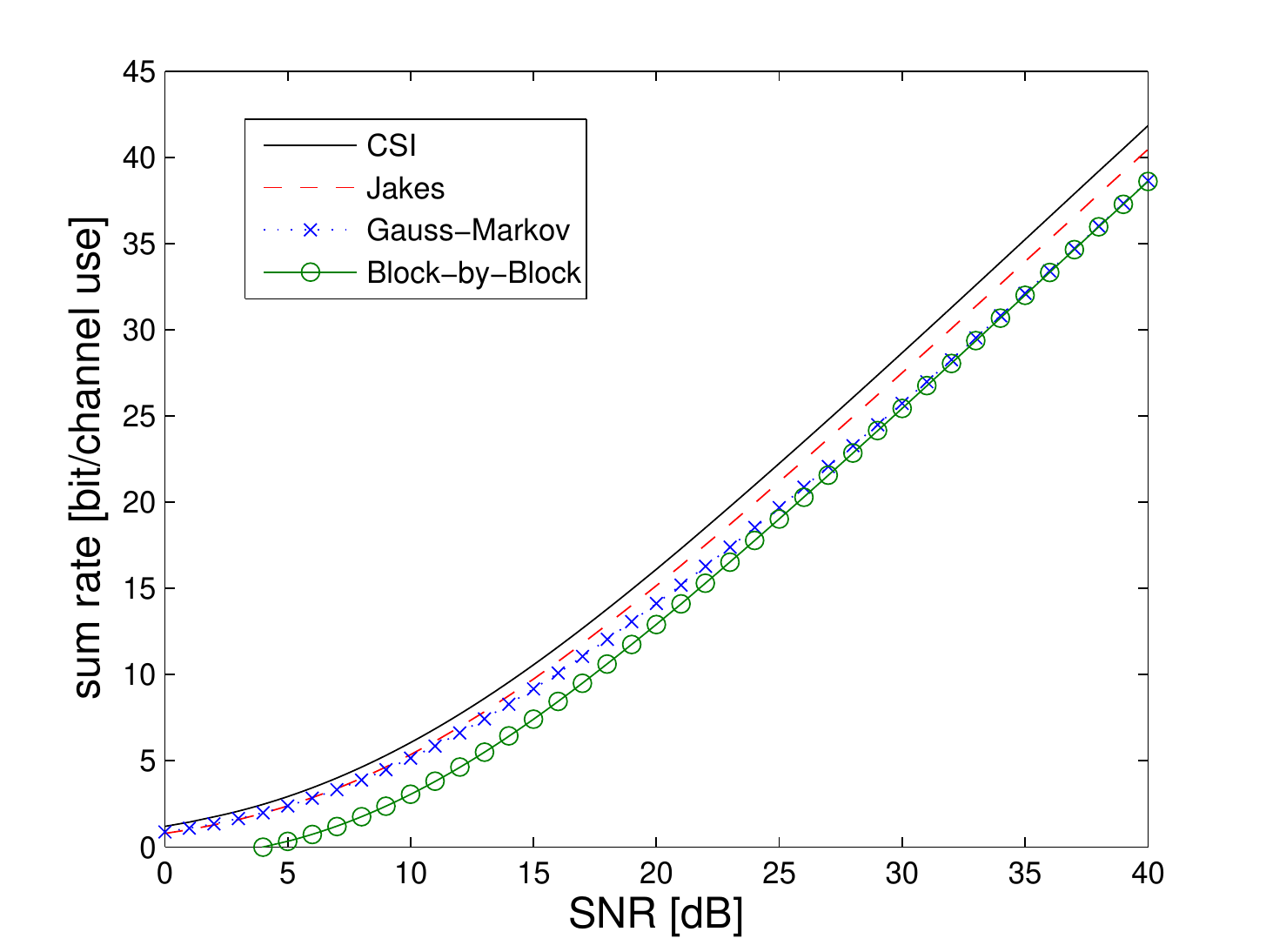}
\vspace{-0.5cm} \caption{Achievable rate lower bounds with optimal
filtering for the Jake's and Gauss-Markov models for $M=4$ and $v =
10$ km/hr ($F=0.0185$ and $r=0.9966$).  Also shown are the rates
with perfect CSI and with block-by-block estimation.}
\label{tput_snr_filtering}
\end{center}
\end{figure}

In Fig. \ref{tput_snr_filtering} the achievable rate lower bound with delay-free
feedback ($d=0$) and optimal filtering is plotted versus SNR for the Jakes and
Gauss-Markov models, for $M=4$, $v = 10$ km/hr ($F=0.0185$ and $r=0.9966$), and $\beta_1 = 1$.
Filtering is seen to provide an advantage with respect to block-by-block estimation
for a wide range of SNR's.  For the Gauss-Markov model this advantage vanishes
around 30 dB, whereas for Jakes' model this advantage persists far beyond the range
of this plot.

\begin{figure}[t]
\begin{center}
\includegraphics[width=10cm]{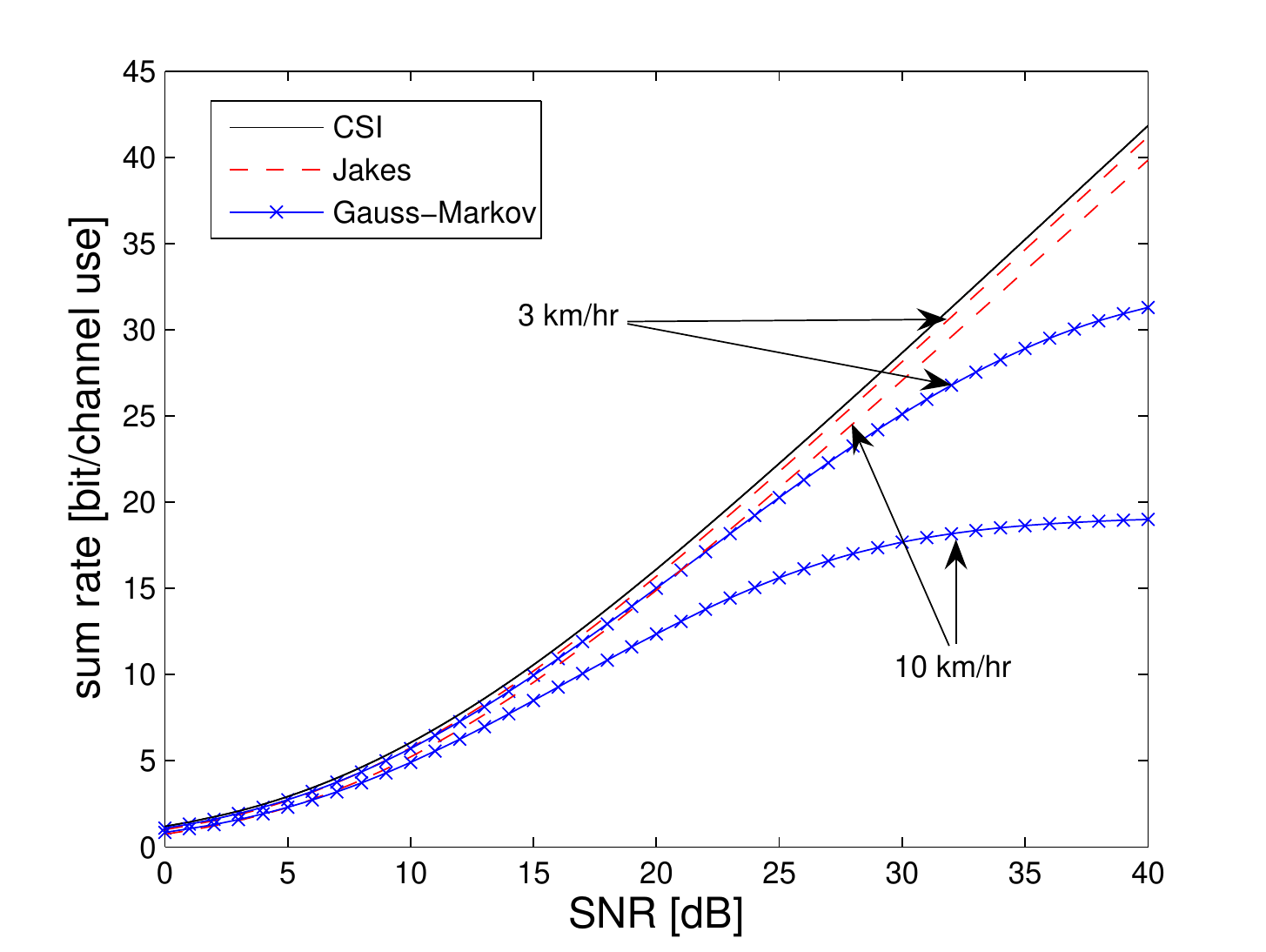}
\vspace{-0.5cm} \caption{Achievable rate lower bounds with optimal
one-step prediction for the Jake's and Gauss-Markov models for
$M=4$.} \label{tput_snr_predict}
\end{center}
\end{figure}

\begin{figure}[t]
\begin{center}
\includegraphics[width=10cm]{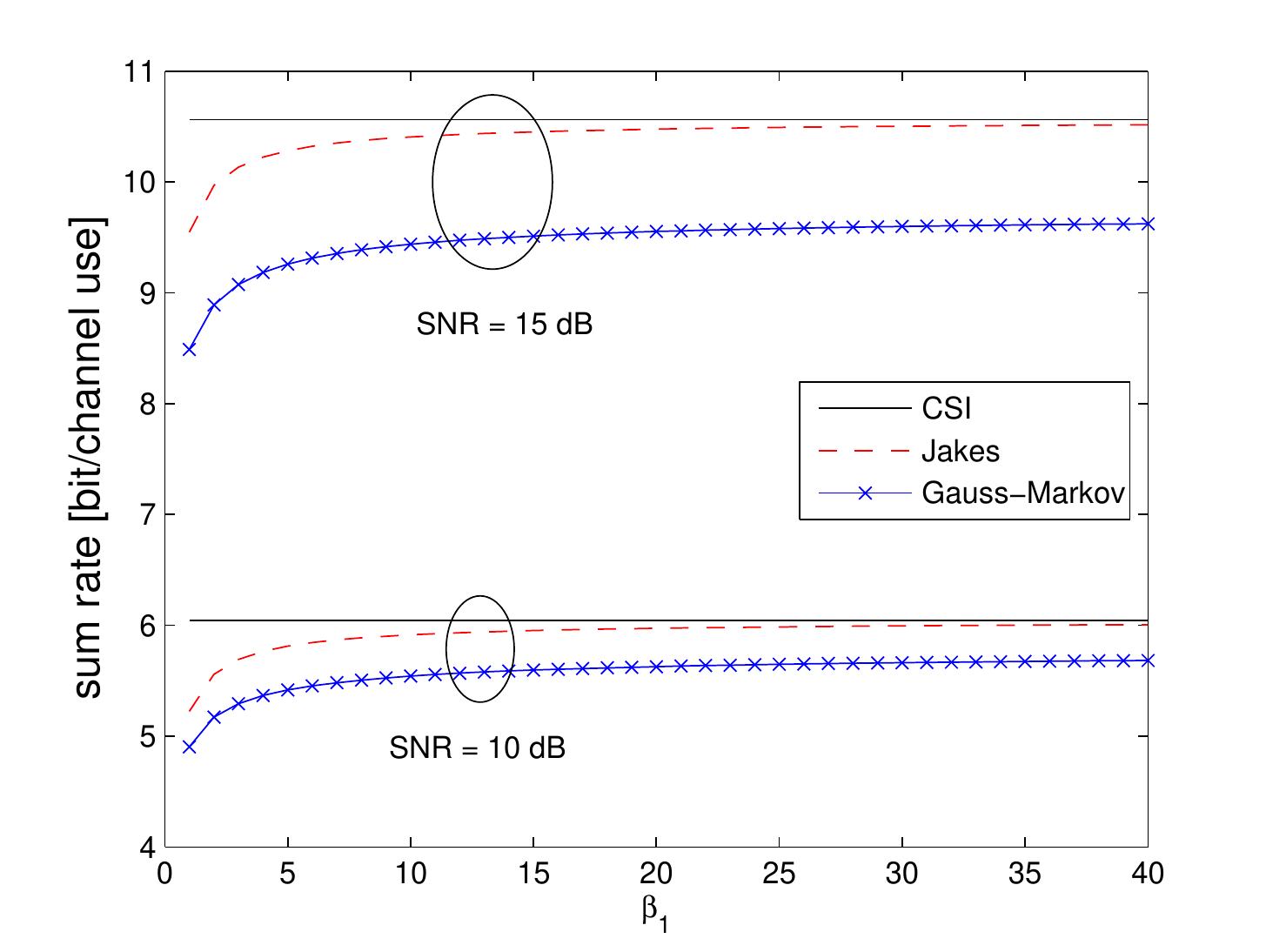}
\vspace{-0.5cm} \caption{Achievable rate lower bounds with optimal
one-step prediction versus $\beta_1$ for $M=4$ and $v = 10$ km/hr.}
\label{tput_beta1_predict}
\end{center}
\end{figure}

Using the same parameters, in Fig. \ref{tput_snr_predict} we plot the lower bound for
one-step prediction ($d=1$) versus SNR for $v = 3$ and $10$ km/hr ($F=0.0056$ and $F=0.0185$).
This figure illustrates the contrast between Doppler and regular processes: for
Jakes' model the achieved rate is quite close to the perfect channel state information
rate (although a slight loss in multiplexing gain is evident), whereas the rate for the
Gauss-Markov model saturates at sufficiently high SNR due to the unpredictability inherent to the model.
To further emphasize the difference in behavior, in Fig. \ref{tput_beta1_predict} we plot the lower bound for one-step
prediction ($d=1$) versus $\beta_1$, the number of common training symbols per block, for $P/N_0 = 10$ and $15$ dB
and $v = 10$ km/hr.  As $\beta_1$ increases (and thus the observation noise decreases)
the rate for Jakes' model converges to the ideal case.
On the other hand, the rate for the Gauss-Markov model saturates at a rate strictly
smaller than the ideal channel state information rate because there is strictly positive prediction
error even if noiseless past observations (i.e., $\beta_1 \rightarrow \infty$) are provided.

In conclusion, the most noteworthy result of this analysis is that under common fading models (Doppler processes),
both analog and digital feedback scheme achieves a potentially high multiplexing gain even with realistic,
noisy and delayed feedback.

\section{Conclusions} \label{conclusions.sect}

This paper presents a comprehensive and rigorous analysis of the achievable performance of ZF beamforming under
pilot-based channel estimation and explicit channel state feedback.
We considered what we believe are the most relevant system aspects.
In particular, the often neglected effect of explicit channel estimation at the UTs is taken into account,
including both common training and dedicated training phases.
As for the feedback, our closed-form bounds allow for a detailed comparison of analog and digital feedback
schemes, including the effects of the MIMO-MAC fading channel, of digital feedback decoding errors,
and of feedback delay.

Our results build on prior work, but generalize many results and
models. We have focused on the case of FDD, but our results easily
extend to TDD systems with channel reciprocity. It is perhaps
important to point out here that our results show that, even in the
case of FDD, a system with explicit CSIT feedback can be
implemented, where the number of training and feedback channel uses
scales linearly with the number of BS antennas, and eventually with
the downlink throughput.

The throughput of the system analyzed here can be
improved via the use of combined beamforming and user
selection/scheduling.
Simulation results show that a system with $K = 10$ and $M = 4$,
with a greedy scheduling as proposed in \cite{DS05,mari-jsac},
achieves a very small gap with respect to the optimal dirty-paper
coding and perfect CSIT case with the same parameters.
Although a clean closed-form analytical characterization of a system with beamforming
and user selection based on imperfect channel state information
appears to be difficult, recent results  \cite{yoo2007jsac,RavindranJindal07}
indicate that the dependence on CSIT quality when user selection is performed is roughly the same as the
equal-power/no selection scenario analyzed here.

We would like to conclude by noticing that some practically relevant
extensions of the present work have been presented (by the same authors and by others)
since the submission of this paper.  In particular, the rate gap analysis was extended to the
very relevant case of MIMO OFDM with frequency-correlated fading in \cite{hooman-TCOM},
the optimal allocation of training and feedback resources is considered in \cite{kobayashi2008mta,kobayashi2009itw},
explicit coding schemes for the CSIT digital feedback MIMO-MAC channel are presented in \cite{raj-ISIT09},
and comparisons between single-user and multi-user MIMO
(based on the bounds developed here and related approximations)
are performed in \cite{zhang_heath}.


\appendices

\section{Proof of Theorem \ref{mutual-info-bound.theorem}} \label{proof_thm1}

The proof is closely inspired by that of Lemma B.0.1 of \cite{lapidoth2002fcp}.
First, notice that since $\widehat{a}_{k,k}$ is a function of $\Rc_k$, by the data-processing inequality we have that
\[ I(u_k; y_k, \Rc_k) \geq I(u_k; y_k, \widehat{a}_{k,k}) \]
Then, because $ I(u_k; y_k, \widehat{a}_{k,k}) = h(u_k) - h(u_k \vert y_k,\widehat{a}_{k,k})$
and $h(u_k) = \log \LB\pi e\frac{P}{M}\RB$, a lower bound on mutual information
is derived by upper bounding $h(u_k \vert y_k,\widehat{a}_{k,k})$ as follows:
\begin{eqnarray} \label{h-bound}
h(u_k \vert y_k,\widehat{a}_{k,k}) & \mathop{=}\limits^{\rm (a)} & h(u_k - \alpha\ y_k \vert y_k, \widehat{a}_{k,k}) \nonumber\\
& \mathop{\leq}\limits^{\rm (b)} & h(u_k - \alpha\ y_k \vert \widehat{a}_{k,k}) \nonumber\\
& \mathop{\leq}\limits^{\rm (c)} & \EE \left [ \log \left ( \pi e \cdot
\EE \left [ | u_k - \alpha\ y_k|^2  | \widehat{a}_{k,k} \right ] \right ) \right ]
\end{eqnarray}
where ${\rm (a)}$ holds for any deterministic function $\alpha$ of $y_k$ and $\widehat{a}_{k,k}$,
${\rm (b)}$ follows from the fact that conditioning reduces entropy and
${\rm (c)}$ follows by the fact that differential entropy is maximized by a Gaussian RV with
the same second moment.
Substituting \eq{train-phase-2} in \eq{receiver-k} we have
\begin{equation} \label{receiver-k-2}
y_k = (\widehat{a}_{k,k} + f_k) u_k + I_k + z_k
\end{equation}
where $\widehat{a}_{k,k} u_k$ and $f_k u_k + I_k + z_k$ are
\emph{uncorrelated} and zero-mean, even if we condition on
$\widehat{a}_{k,k}$, because  $\widehat{a}_{k,k}, f_k, u_1,\ldots,
u_K, z_k$ are independent, zero-mean Gaussian's.    Thus, we have
\begin{equation} \label{y_k-var}
\EE\LSB|y_k|^2\ \vert \widehat{a}_{k,k}\RSB = |\widehat{a}_{k,k}|^2
\EE[|u_k|^2] + \sigma^2_2\ \EE[|u_k|^2]  + \EE\LSB
|I_k|^2|\widehat{a}_{k,k}\RSB + N_0,
\end{equation}
Choosing $\alpha$ that minimizes $\EE\LSB |u_k - \alpha\ y_k |^2 |
\widehat{a}_{k,k}\RSB$ tightens the bound. This corresponds to
setting $\alpha\ y_k$ equal to the linear MMSE estimate of $u_k$
given $y_k$ and $\widehat{a}_{k,k}$, i.e.,
\begin{equation} \label{alpha-LMMSE}
\alpha = \frac{\EE\LSB u_k y_k^*\ \vert\ \widehat{a}_{k,k}\RSB}{\EE\LSB|y_k|^2\ \vert\ \widehat{a}_{k,k}\RSB}
= \frac{\EE[|u_k|^2] \widehat{a}_{k,k}^*}{\EE\LSB|y_k|^2\ \vert\ \widehat{a}_{k,k}\RSB}
\end{equation}
Using \eq{y_k-var}, the corresponding MMSE is given by
\begin{eqnarray}
\label{succhia}
\EE \LSB |u_k - \alpha\ y_k|^2 | \widehat{a}_{k,k}\RSB & = & \EE\LSB |u_k|^2 \RSB \LB 1 - \frac{\EE[|u_k|^2] |\widehat{a}_{k,k}|^2 }{\EE\LSB|y_k|^2\ \vert\ \widehat{a}_{k,k} \RSB} \RB \\
\label{alpha-LMMSE-error}
& = & \frac{P}{M} \frac{1 + \sigma_2^2 \frac{P}{N_0 M} + \EE\LSB |I_k|^2|\widehat{a}_{k,k}\RSB/N_0}{|\widehat{a}_{k,k}|^2
\frac{P}{N_0 M} + 1 + \sigma_2^2 \frac{P}{N_0 M} + \EE\LSB |I_k|^2|\widehat{a}_{k,k}\RSB/N_0}
\end{eqnarray}
Replacing \eq{alpha-LMMSE-error} into \eq{h-bound} and using $h(u_k) = \log \LB\pi e\frac{P}{M}\RB$,
we obtain (\ref{mutual-info-bound}).

\section{Proof of Theorem \ref{rate-gap-bound.theorem}} \label{proof_thm2}

Using the lower bound on $R_k$ from Theorem \ref{mutual-info-bound.theorem}
we have:
\begin{eqnarray}
\Delta R
& \leq &  \EE \left[ \log \left(1 +  \frac{|\hv^\herm \vv_k|^2  P}{N_0M} \right) \right] -  \EE \left[ \log\left( 1 + \frac{|\widehat{a}_{k,k}|^2 P/(N_0 M)} {1 + \sigma_2^2 P/(N_0 M) + \EE\LSB |I_k|^2|\widehat{a}_{k,k}\RSB/N_0}\right) \right]\nonumber \\
& \mathop{\leq}\limits^{\rm (a)} & \EE \left[ \log \left(1 +  \frac{|\hv^\herm\vv_k|^2  P}{N_0M} \right) \right]
- \EE \left[ \log\left( 1 + \frac{P}{N_0M} \LB |\widehat{a}_{k,k}|^2  + \sigma_2^2 \RB \right) \right] \nonumber\\
& & \qquad\qquad\qquad\qquad\qquad\ \ +\ \EE \left[ \log\left( 1 + \sigma_2^2 \frac{P}{N_0M} + \frac{\EE\LSB |I_k|^2|\widehat{a}_{k,k}\RSB}{N_0}\right) \right] \nonumber\\
& \mathop{\leq}\limits^{\rm (b)} & \EE \left[ \log\left( 1 + \sigma_2^2 \frac{P}{N_0M} + \frac{\EE\LSB |I_k|^2|\widehat{a}_{k,k}\RSB}{N_0}\right) \right] \label{intermed-bound}\\
& \mathop{\leq}\limits^{\rm (c)} & \log\left( 1 + \sigma_2^2 \frac{P}{N_0M} + \frac{\EE[|I_k|^2]}{N_0}\right)
\end{eqnarray}
where ${\rm (a)}$ follows by dropping the non-negative term $\EE\LSB |I_k|^2|\widehat{a}_{k,k}\RSB/N_0$.
Using the fact that $\hv_k$ is spatially white and $\vv_k$ is selected independent of
$\hv_k$ (by the ZF procedure), it follows that $\hv_k^\herm \vv_k$
is $\sim \Cc\Nc(0,1)$ and $\widehat{a}_{k,k} \sim
\Cc\Nc(0,1-\sigma_2^2)$. Direct application of Lemma
\ref{littlelemma.lemma}, which is provided below, with $A = P/(N_0M)$, $\lambda = \sigma_2^2$
and $X=|\hv_k^\herm \vv_k|^2$,
thus proves
${\rm (b)}$. Finally, ${\rm (c)}$ follows from the concavity of
$\log(\cdot)$ and Jensen's inequality.

\begin{lemma} \label{littlelemma.lemma}
If $X$ is a non-negative random variable with $\EE[X] = 1$,
for any  $A>0$ and  any $0 \leq \lambda \leq 1$:
\begin{equation} \label{littlelemma.eqn}
\EE\LSB \log\LB 1 + X A \RB \RSB \leq \EE\LSB \log\LB 1 + \LB \lambda + (1-\lambda)X \RB A \RB \RSB.
\end{equation}
\end{lemma}

{\bf Proof:}
For all $0 \leq z \leq 1$, define the function
\begin{equation}
\psi(z) = \EE \LSB\log \LB 1 + z A + (1-z) X A \RB\RSB.
\end{equation}
Then \eq{littlelemma.eqn} is equivalent to the inequality $\psi(0) \leq \psi(\lambda)$.
By the concavity of $\log(\cdot)$ and Jensen's inequality we have
\begin{eqnarray}
\psi(z) & \leq & \log \LB 1 + z A + (1-z) \EE\LSB X\RSB A \RB
=  \psi(1).
\end{eqnarray}
In particular, $\psi(0) \leq \psi(1)$. Moreover, $\psi(z)$ is an expectation of the composition of a concave
function and a linear function of $z$, and is hence concave \cite{boyd2004co}.
Thus, the concave function $\psi(z)$ for $z \in [0, 1]$ lies above the line joining the points $(0, \psi(0))$ and $(1, \psi(1))$. Hence, we have $\psi(0) \leq \psi(\lambda)$ for $\lambda \in [0, 1]$, which proves \eq{littlelemma.eqn}.
\QED

\section{Proof of Theorem \ref{thm:analog_rate_gap}} \label{proof_thm4}

Using (\ref{eq-rategap_simple}), to compute $\overline{\Delta R}^\textsc{AF}$
we only need to find $\EE\LSB |\hv_k^\herm\widehat{\vv}_j|^2\RSB$:
\begin{eqnarray}
\EE\LSB |\hv_k^\herm\widehat{\vv}_j|^2\RSB
& \mathop{=}\limits^{\textrm{(a)}} & \EE\LSB |\widehat{\hv}_k^\herm\widehat{\vv}_j
+ {\bf \ev}_k^\herm\widehat{\vv}_j|^2\RSB  \nonumber\\
& \mathop{=}\limits^{\textrm{(b)}} & \EE\LSB |{\bf \ev}_k^\herm\widehat{\vv}_j|^2\RSB \nonumber\\
& \mathop{=}\limits^{\textrm{(c)}} & \EE\LSB \widehat{\vv}_j^\herm\EE[{\bf \ev}_k {\bf \ev}_k^\herm] \widehat{\vv}_j\RSB \nonumber\\
& \mathop{=}\limits^{\textrm{(d)}} & \sigma_e^2 \label{anlogfb-var}
\end{eqnarray}
where ${\rm  (a)}$ follows from (\ref{analogfb-iid}),
 ${\rm  (b)}$ follows from the fact that $\widehat{\hv}_k^\herm\widehat{\vv}_j = 0\ \forall\ j \neq k$ by
naive ZF, ${\rm (c)}$ is obtained from the independence of $\ev_k$ and $\widehat{\vv}_j$
($\widehat{\vv}_j$ is a deterministic function of $\{\widehat{\hv}_i\}_{i\neq j}$),
and (d) follows from $\EE[{\bf \ev}_k {\bf \ev}_k^\herm] = \sigma_e^2 {\bf I}$ and
$\|\widehat{\vv}_j\| = 1$.

\section{Proof of Theorem \ref{thm:digital_rate_gap}} \label{proof_thm5}

To compute the rate gap upper bound, we determine $\EE\LSB |\hv_k^\herm\widehat{\vv}_j|^2\RSB$
by writing the channel in terms of the UT channel estimate (which is quantized)
and the UT estimation error: $\hv_k = \widetilde{\hv}_k + \nv_k$ from (\ref{train-phase-1}).
This yields:
\begin{eqnarray}
\EE\LSB |\hv_k^\herm\widehat{\vv}_j|^2\RSB
& \mathop{=}\limits^{\textrm{(a)}} & \EE \LSB |\widetilde{\hv}_k^\herm\widehat{\vv}_j|^2\RSB + \EE\LSB |{\bf \nv}_k^\herm\widehat{\vv}_j|^2\RSB \nonumber \\
& \mathop{=}\limits^{\textrm{(b)}} & \EE\LSB \|\widetilde{\hv}_k\|^2 \RSB \EE\LSB \frac{|\widetilde{\hv}_k^\herm\widehat{\vv}_j|^2}{\|\widetilde{\hv}_k\|^2}\RSB + \EE\LSB |{\bf \nv}_k^\herm\widehat{\vv}_j|^2\RSB \nonumber\\
& \mathop{=}\limits^{\textrm{(c)}} & \frac{\EE\LSB\|\widetilde{\hv}_k\|^2\RSB}{M-1} \ 2^B \beta \left( 2^B, \frac{M}{M-1}
\right)
+  \EE\LSB \widehat{\vv}_j^\herm\EE[{\bf \nv}_k {\bf \nv}_k^\herm] \widehat{\vv}_j\RSB \nonumber\\
& \mathop{=}\limits^{\textrm{(d)}} &
\frac{M}{M-1} \frac{\beta_1 P}{N_0+\beta_1P}\ 2^B \beta \left( 2^B, \frac{M}{M-1} \right)
+ \sigma_1^2  \label{cross-term_digital}
\end{eqnarray}
where (a) is obtained from the representation $\hv_k = \widetilde{\hv}_k + \nv_k$ and the fact that
$\EE \LSB \widetilde{\hv}_k^\herm \widehat{\vv}_j  \widehat{\vv}_j^\herm \nv_k \RSB = 0$
because $\nv_k$ is zero-mean Gaussian and is independent of
$\widetilde{\hv}_k$ and $\widehat{\vv}_j$,
(b) from the independence of the channel norm and direction of $\widetilde{\hv}_k$,
(c) from \eq{RVQ-bound} and from the property \cite[Lemma 2]{Jindal}
$\EE\LSB\frac{|\widetilde{\hv}_k^\herm\widehat{\vv}_j|^2}{\|\widetilde{\hv}_k\|^2}\RSB
= \frac{1}{M-1} \EE\LSB \sin^2 \left( \widetilde{\hv}_k,  \widehat{\hv}_k \right) \RSB,$
and finally (d) by computing the expected norm of $\widetilde{\hv}_k = \frac{\sqrt{\beta_1 P}}{N_0 + \beta_1 P} \sv_k$
using $\sv_k = \sqrt{\beta_1 P}\ \hv_k + \zv_k$.  The final result follows by
using the above result in the expression (\ref{rate-gap-bound}) for the rate gap.

\section{Proof of Theorem \ref{thm:digital_rate_gap_errors}} \label{proof_thm6}

We first decompose the interference variance term as
\begin{eqnarray}
\EE\LSB |\hv_k^\herm\widehat{\vv}_j|^2\RSB &=& (1 - P_{e,{\rm fb}})
\EE [|\hv_k^\herm\widehat{\vv}_j|^2|\mbox{no fb. errors}] + P_{e,{\rm fb}}
\EE [|\hv_k^\herm\widehat{\vv}_j|^2|\mbox{fb. errors}] \\
&\leq& (1 - P_{e,{\rm fb}})  \frac{M}{M-1} \frac{\beta_1 P}{N_0+\beta_1P}\ 2^B \beta \left( 2^B, \frac{M}{M-1} \right)
+ \sigma_1^2 + P_{e,{\rm fb}},
\end{eqnarray}
where $\EE [|\hv_k^\herm\widehat{\vv}_j|^2|\mbox{no fb. errors}]$ is the same as in the error-free case and is thus given in
(\ref{cross-term_digital}) while for the case of feedback errors we trivially have $\EE [|\hv_k^\herm\widehat{\vv}_j|^2|\mbox{fb. errors}] \leq 1$.
The final result is reached by simply substituting $B = \alpha(M-1) \log_2 \frac{P}{N_0}$ and using
the bound in the beta function (\ref{RVQ-bound}).

\section{Proof of Theorem \ref{thm:analog_mimo_rate_gap}} \label{pf_mimo_mac_analog}

Using the argument from the proof of Theorem \ref{proof_thm4}
(analog FB over AWGN channel), the
expected interference coefficient $\EE\LSB |\hv_k^\herm\widehat{\vv}_j|^2\RSB$ is
is equal to the variance of the channel estimation error.  This quantity must be averaged
over the uplink channel matrix ${\bf A}$, and thus using symmetry and (\ref{ziopino2}), is given by
\begin{eqnarray} \label{ziopino3}
\EE[\sigma_k^2(\Am)] & =& \EE \left [ \frac{1}{L} \trace \left ( \Id - c^2 \Am^\her \left [ \beta_{\rm fb}P \Am\Am^\her +
N_0 \Id \right ]^{-1} \Am \right ) \right ] \nonumber \\
& = & \EE \left [ \frac{1}{L} \sum_{k=1}^L \frac{N_0 + (\beta_{\rm fb} P - c^2)\lambda_k}
{N_0 + \beta_{\rm fb} P \lambda_k} \right ] \nonumber \\
& = & \frac{1}{1 + \beta_1 \frac{P}{N_0}} +
\frac{\beta_1 \frac{P}{N_0}}{1 + \beta_1 \frac{P}{N_0}} {\sf mmse} \left ( \beta_{\rm fb} \frac{P}{N_0} \right )
\end{eqnarray}
where ${\sf mmse}(\rho)$ is defined in (\ref{ziopino4}).

In order to obtain the high SNR result, we first state a closed-form expression for ${\sf mmse}(\rho)$
using well-known results from multivariate statistics (see for example \cite{shin2003cma}):
\begin{equation} \label{mari-result}
{\sf mmse}(\rho) = \frac{e^{1/\rho}}{\rho} \sum_{k=0}^{L-1} \sum_{\ell=0}^k \sum_{m=0}^{2\ell}
{\sf X}_{k,\ell,m} {\rm E_i}(M-L+m+1,1/\rho)
\end{equation}
where the coefficients ${\sf X}_{k,\ell,m}$ are given by
\[ {\sf X}_{k,\ell,m} = \frac{(-1)^m (2\ell)! (M-L+m)!}{L 2^{2k-m}\ell ! m! (M-L+\ell)!} {{2(k-\ell)} \choose {k-\ell}} {{2(M-L+\ell)} \choose {2\ell-m}} \]
Based upon this we can characterize the asymptotic behavior of the product $\rho \; {\sf mmse}(\rho)$ for $\rho \rightarrow \infty$.
Using the asymptotic expansion of $e^{1/\rho} {\rm E_i}(n,1/\rho)$, we have
\begin{equation} \label{ziopino5}
\rho \ {\sf mmse}(\rho) = \left \{
\begin{array}{ll}
\frac{1}{M-L} + o(1) & \mbox{for} \;\; L < M \\
-\gamma + \log \rho + \sum_{k=0}^{L-1} \sum_{\ell=0}^k \sum_{m=1}^{2\ell} \frac{{\sf X}_{k,\ell,m}}{m} + o(1)
& \mbox{for} \;\; L = M \end{array} \right .
\end{equation}
where we used the facts:
\begin{eqnarray}
{\rm E_i}(1,1/\rho) e^{1/\rho} & = & - \gamma + \log \rho + o(1), \;\;\; \rho \rightarrow \infty \\
{\rm E_i}(n,1/\rho) e^{1/\rho} & = & \frac{1}{n-1} + o(1), \;\;\; \mbox{for} \; n > 1, \; \rho \rightarrow \infty \\
\sum_{k=0}^{L-1} \sum_{\ell=0}^k \sum_{m=0}^{2\ell} \frac{{\sf X}_{k,\ell,m}}{M-L+m} & = & \frac{1}{M-L},  \;\;\; \mbox{for}\; L < M \\
\sum_{k=0}^{L-1} \sum_{\ell=0}^k {\sf X}_{k,\ell,0} & = & 1, \;\;\; \mbox{for} \; L = M
\end{eqnarray}

\section{Proof of Theorem \ref{thm:geniegapLM}}\label{proof_geniebound}

We can lower bound the genie-aided rate  of Theorem \ref{genie-mutual-info-bound.theorem} as follows.
\begin{eqnarray*}
 I(u_k; y_k, \Ac_k) & = & \EE \left[ \log\left( 1 + \frac{|a_{k,k}|^2 P/(N_0 M)} {1 + \sum_{j \neq k} |a_{k,j}|^2 P/(N_0 M)} \right) \right] \nonumber\\
& = & \EE \left[ \log\left( 1 + \sum_{j} |a_{k,j}|^2 P/(N_0 M) \right) \right]  - \EE \left[ \log\left( 1 + \sum_{j \neq k} |a_{k,j}|^2 P/(N_0 M) \right) \right]\\
&\overset{\mathrm{(a)}}{\geq} & R_k^\textsc{ZF}  - \EE \left[ \log\left( 1 + \sum_{j \neq k} |a_{k,j}|^2 \frac{P}{N_0 M} \right) \right] \\
& \overset{\mathrm{(b)}}{\geq} & R_k^\textsc{ZF}  - \EE  \left[ \log\left( 1 + \sum_{j\neq k}  \EE[ |a_{k,j}|^2 \vert \Am] \frac{P}{N_0 M} \right) \right]\\
& \overset{\mathrm{(c)}}{=} & R_k^\textsc{ZF}  - \EE  \left[ \log\left( 1 + \frac{P}{N_0 M}\sigma_k^2(\Am) \right) \right]
\end{eqnarray*}
where (a) follows by dropping the non-negative terms and (b)
 follows by conditioning with respect to the uplink channel matrix $\Am$ and then
applying Jensen's inequality in the inner conditional expectation, (c) follows by noticing $\EE[|I_k|^2|\Am]= (M-1)P\sigma_e^2(\Am)$ where $\sigma_k^2(\Am)$ is defined in (\ref{ziopino2}).
Then, we obtain an upper bound of for the gap between the ideal ZF rate and the genie-aided rate given by
\begin{eqnarray}\nonumber
R_k^\textsc{ZF} - I(u_k;y_k,\Ac_k) & \leq & \EE \left [ \log\left( 1 + \frac{P}{N_0} \frac{M-1}{M}  \sigma_k^2(\Am) \right) \right]\\  \nonumber
& \stackrel{\rm (a)}{=} & \frac{1}{M} \sum_{k=1}^M \EE \left [ \log\left( 1 + \frac{P}{N_0} \frac{M-1}{M}  \sigma_k^2(\Am) \right) \right]  \\ \nonumber
& \leq & \EE \left [ \log\left( 1 + \frac{P}{N_0} \frac{M-1}{M}  \frac{1}{M} \sum_{k=1}^M \sigma_k^2(\Am) \right) \right]  \\ \nonumber
& \stackrel{\rm (b)}{=} &
\EE \left [ \log\left( 1 + \frac{P}{N_0} \frac{M-1}{M} \left ( \frac{1}{1 + \beta_1 \frac{P}{N_0}} +
\frac{\beta_1 \frac{P}{N_0}}{1 + \beta_1 \frac{P}{N_0}} \frac{1}{M} \sum_{k=1}^M \frac{1}{1 + \beta_{\rm fb} \frac{P}{N_0} \lambda_k}
  \right ) \right ) \right ] \\ \label{fullmmse3} 
& \leq  &
\EE \left [ \log\left( 1 + \frac{M-1}{M} \left ( \frac{1}{\beta_1} +
\frac{\frac{P}{N_0}}{1 + \beta_{\rm fb} \frac{P}{N_0} \lambda_{\min} }
  \right ) \right ) \right ]
\end{eqnarray}
where (a) follows because the term $\EE \left [ \log\left( 1 + \frac{P}{N_0} \frac{M-1}{M}  \sigma_k^2(\Am) \right) \right]$ is independent of $k$ due to the symmetry over $k$, (b) follows by using the same derivation that leads to (\ref{ziopino3}) and (\ref{ziopino4}), and the last line follows by monotonicity of the log, where $\lambda_{\min}$ denotes the minimum
eigenvalue of $\Am^\herm \Am$.

Our goal is to show that the term in the last line of (\ref{fullmmse3}) is bounded.
To this purpose, we write the last line of (\ref{fullmmse3}) as the sum of three terms,
\begin{eqnarray}
&
\log \left (1 + \frac{M-1}{M\beta_1} + \frac{M-1}{M} \frac{P}{N_0} \right ) +
\EE \left [ \log \left ( 1 + \frac{\left (1 + \frac{M-1}{M\beta_1} \right ) \frac{\beta_{\rm fb} P}{N_0} }{1 + \frac{M-1}{M} \left (\frac{1}{\beta_1} +
\frac{P}{N_0} \right ) } \lambda_{\min} \right ) \right ]
- \EE\left [ \log \left (1 + \beta_{\rm fb} \frac{P}{N_0} \lambda_{\min} \right ) \right ] & \nonumber \\
& & \label{terms}
\end{eqnarray}
For $\Am$ $M \times M$, complex Gaussian with i.i.d. zero-mean components,
it is well-known that $\lambda_{\min}$ is chi-squared
with 2 degrees of freedom and mean 1 \cite{edelman-thesis}.
Hence, the third term in (\ref{terms}) yields
\[ \EE\left [ \log \left (1 + \beta_{\rm fb} \frac{P}{N_0} \lambda_{\min} \right ) \right ] = e^{\frac{N_0}{\beta_{\rm fb} P}} {\rm E_i}\left (1 ,
\frac{N_0}{\beta_{\rm fb} P} \right ) = -\gamma +\log \frac{\beta_{\rm fb} P}{N_0} + o(1) \]
The second term in (\ref{terms}) is bounded by a constant, independent of $P/N_0$, and finally
the first term in (\ref{terms}), for high SNR, can be written as  $\log \frac{P}{N_0} + O(1)$.
It follows that the $\log(P/N_0)$ terms in the first and the third terms of the
the upper bound cancel, so that (\ref{terms}) is bounded. This establishes the result.


\section{Genie-Aided Upper Bound for Regular Processes with Delayed Feedback}  \label{app-bounded}
We show that the genie-aided upper bound of Theorem
\ref{genie-mutual-info-bound.theorem}, is uniformly bounded for any
SNR when the noiseless prediction error is positive. For analytical
simplicity, we assume perfect common training and perfect (delayed)
feedback. Hence, the only source of ``noise'' in the CSIT is due to
the prediction error. We can write $\hv_k(t) = \widetilde{\hv}_k(t)
+ \nv_k(t)$, where $\widetilde{\hv}_k(t)$ is the one-step prediction
of $\hv_k(t)$ from its (noiseless) past, and $\nv_k(t)$ is the
prediction error. From what was stated earlier, we have that $\hv_k(t),
\widetilde{\hv}_k(t)$ and $\nv_k(t)$ are jointly complex Gaussian,
i.i.d. in the spatial domain, with mean zero and variance per
component equal to $1, 1 - \epsilon_1(0)$ and $\epsilon_1(0)$,
respectively. It is useful to write the error as $\nv_k(t) =
\sqrt{\epsilon_1(0)} \Deltam(t)$, where $\Deltam(t) \sim
\Cc\Nc(\zerov, \Id)$. From \eq{genie-mutual-info-bound}, the
genie-aided upper bound  is given by \begin{equation} \label{cacane}
R_k \leq \EE \LSB \log \LB 1 + \frac{P |\hv_k^\herm(t)
\hat{\vv}_k(t) |^2}{N_0M  +  P \sum_{j \neq k} |\hv_k^\herm(t)
\widehat{\vv}_j(t) |^2} \RB \RSB \nonumber \\ \end{equation} where
$\hat{\vv}_j(t)$ is orthogonal to $\widetilde{\hv}_k(t)$. Using the
fact that the upper bound is non-decreasing in $P/N_0$, we let
$P/N_0 \rightarrow \infty$ in (\ref{cacane}) and obtain
\begin{eqnarray} \label{ceiling} R_k & \leq & \EE \LSB \log \LB
|\hv_k^\herm(t) \widehat{\vv}_k(t)|^2 + \sum_{j \neq k}
|\hv_k^\herm(t)\widehat{\vv}_j(t)|^2 \RB \RSB - \EE \LSB \log \LB
\sum_{j \neq k} |\hv_k^\herm(t)\widehat{\vv}_j(t)|^2 \RB \RSB
\nonumber \\ & \mathop{\leq}\limits^{(a)} & \log \LB 1 +
\epsilon_1(0)(M-1) \RB - \EE \LSB \log \LB \epsilon_1(0) \sum_{j
\neq k} |\Deltam_k^\herm(t) \widehat{\vv}_j(t)|^2 \RB \RSB
\nonumber \\ & \mathop{=}\limits^{(b)} & \log \LB
\frac{1}{\epsilon_1(0)} + M-1 \RB - \EE [ \log (|\Deltam_k(t)|^2) ]
- \EE \LSB \log \LB \sum_{j \neq k}
\frac{|\Deltam_k^\herm(t)\widehat{\vv}_j(t)|^2}{|\Deltam_k(t)|^2}
\RB \RSB \nonumber\\ & \mathop{=}\limits^{(c)} & \log \LB
\frac{1}{\epsilon_1(0)} + M-1 \RB - \psi(M)  + \frac{1}{2M-1} +
\frac{1}{2M-2} \end{eqnarray} where (a) follows by applying Jensen's
inequality to the first term and noticing  that both
$\hv_k^\herm(t)\widehat{\vv}_k(t)$ and
$\Deltam_k^\herm(t)\widehat{\vv}_j(t)$ are $\sim \Cc\Nc(0,1)$, (b)
follows by expressing $|\Deltam_k^\herm(t)\widehat{\vv}_j(t)|^2
=|\Deltam_k^\herm(t)|^2
\frac{|\Deltam_k^\herm(t)\widehat{\vv}_j(t)|^2
}{|\Deltam_k^\herm(t)|^2}$, (c) is obtained by noticing that
$|\Deltam_k(\tau)|^2$ is chi-square distributed with $2M$ degrees of
freedom and that $\sum_{j \neq k}
\frac{|\Deltam_k^\herm(t)\widehat{\vv}_j(t)|^2} {|\Deltam_k(t)|^2}$
is beta distributed with parameters $(M-1, 1)$, and finally
$\psi(M)$ is the Euler-Digamma function. \hfill $\lozenge$

\centerline{Acknowledgment}
The work of G. Caire was partially supported by NSF Grant CCF-0635326.

\newpage


\end{document}